\documentclass[12pt]{article}

\pdfoutput=1

\usepackage[top=80pt,bottom=85pt,left=85pt,right=85pt]{geometry}
\usepackage{amssymb}
\usepackage{amsmath}
\usepackage{setspace}
\usepackage{sectsty}
\usepackage{cite}
\usepackage{bbm}
\usepackage[utf8]{inputenc}
\usepackage[T1]{fontenc}
\usepackage{comment}
\usepackage[english]{babel}
\usepackage{afterpage}
\usepackage{bbold}
\usepackage{mathdots}
\usepackage{physics}
\usepackage{multirow}
\usepackage[dvipsnames]{xcolor}

\usepackage{graphicx,subcaption}
\usepackage{setspace}
\usepackage{float}
\usepackage{booktabs}
\usepackage[debug,pageanchor=false]{hyperref}
\definecolor{link}{rgb}{.8,.15,.1}
\definecolor{pigment}{rgb}{0.36, 0.54, 0.66}
\definecolor{pigment2}{rgb}{0.19, 0.55, 0.91}
\definecolor{pigment3}{rgb}{0.2, 0.2, 0.6}
\definecolor{light-gray}{gray}{0.75}
\hypersetup{colorlinks=true,linkcolor=link,citecolor=link,urlcolor=link,linktocpage}

\usepackage{array,multirow,booktabs,longtable}
\usepackage{mathrsfs}
\usepackage{tikz-cd} 
\usetikzlibrary{backgrounds, arrows,calc,shapes,decorations.pathreplacing, decorations.markings, automata,positioning}
\tikzset{%
  >={Latex[width=2mm,length=2mm]},
            base/.style = {rectangle, rounded corners, draw=black,
                           minimum width=4cm, minimum heigwht=1cm,
                           text centered, font=\sffamily},
  activityStarts/.style = {base, fill=orange!15},
       startstop/.style = {base, fill=orange!15},
    activityRuns/.style = {base, fill=orange!15},
         process/.style = {base, minimum width=2.5cm, fill=orange!15,
                           font=\ttfamily},
}

\newcommand{\red}[1]{}

\tikzset{
        cvertex/.style={circle,draw=black,inner sep=1pt,outer sep=3pt},
        vertex/.style={circle,fill=black,inner sep=1pt,outer sep=3pt},
        star/.style={circle,fill=yellow,inner sep=0.75pt,outer sep=0.75pt},
        tvertex/.style={inner sep=1pt,font=\scriptsize},
        gap/.style={inner sep=0.5pt,fill=white}}

\tikzstyle{mybox} = [draw=black, fill=blue!10, very thick,
    rectangle, rounded corners, inner sep=10pt, inner ysep=20pt]
\tikzstyle{boxtitle} =[fill=blue!50, text=white,rectangle,rounded corners]

\newcommand\scalemath[2]{\scalebox{#1}{\mbox{\ensuremath{\displaystyle #2}}}}
\newcolumntype{C}{>{\hfil$}p{0.65cm}<{$\hfil}}
\newcolumntype{P}{>{\hfil$}p{6.4cm}<{$\hfil}}
\newcolumntype{L}{>{\hfil$}p{1.9cm}<{$\hfil}}
\newcolumntype{R}{>{\hfil$}p{0.8cm}<{$\hfil}}

\newcommand{\todo}[1]{}
\renewcommand{\todo}[1]{{\color{red} TODO: {#1}}}
\renewcommand{\red}[1]{{\color{red} {#1}}}

\newcommand{\be}{\begin{equation}}  
\newcommand{\ee}{\end{equation}}  
\newcommand{\bea}{\begin{align}}
\newcommand{\eea}{\end{align}}
\newcommand{\bp}{\begin{bmatrix*}[r]}  
\newcommand{\bpp}{\begin{bmatrix}}  
\newcommand{\epp}{\end{bmatrix}}  
\newcommand{\bcd}{\begin{center}
\begin{tikzcd}}
\newcommand{\ecd}{\end{tikzcd} \end{center}}
\newcommand{\bm}{\begin{pmatrix}}  
\newcommand{\eem}{\end{pmatrix}}

\makeatletter
\@addtoreset{equation}{section}
\makeatother

\begin{document}


\begin{titlepage}

\begin{center}

\vskip .3in \noindent

{\Large \bf{Higgs Branches of rank-0 5d theories from \\  
M-theory on $\boldsymbol{(A_j,A_l)}$ and $\boldsymbol{(A_k,D_n)}$ singularities}}

\bigskip\bigskip

Mario De Marco$^{a}$, Andrea Sangiovanni$^{b}$ \\

\bigskip


\bigskip
{\footnotesize
 \it

$^a$ SISSA and INFN, Via Bonomea 265, I-34136 Trieste, Italy\\
\vspace{.25cm}
$^b$ Dipartimento di Fisica, Universit\`a di Trieste, Strada Costiera 11, I-34151 Trieste, Italy \\
and INFN, Sezione di Trieste, Via Valerio 2, I-34127 Trieste, Italy	
}

\vskip .5cm
{\scriptsize \tt  mdemarco at sissa dot it \hspace{2cm} andrea dot sangiovanni at phd dot units dot it} \\

\vskip 1cm
     	{\bf Abstract }
\vskip .1in
\end{center}
We study the dynamics of M-theory on isolated non-toric Calabi-Yau threefold singularities of type $(A_j,A_l)$ and $(A_k,D_n)$, engineering five-dimensional rank-zero SCFTs. Our approach consists in mapping these backgrounds to type IIA string theory with D6 branes at angles and O$6^{-}$ planes, computing the five-dimensional open string modes located at the brane intersections. This permits us to characterize the Higgs Branches of these theories as algebraic varieties, determine the flavour and gauge group and inspect subtleties related to T-branes. Our methods apply for all the considered singularities: we give a closed formula for the $(A_j,A_l)$ Higgs Branches, and tables for the Higgs Branches of the $(A_k,D_n)$ series.

\noindent

\vfill
\eject

\end{titlepage}


\tableofcontents


\section{Introduction}
\label{sec:introduction}

Conformal field theories appear ubiquitously in physics. From the point of
view of gauge theories, conformal invariance seems to be a feature
typically arising at strong coupling regimes, making the study of
conformal field theories particularly interesting. Recently \cite{Zotto:bj,Bhardwaj:2018vuu,Cabrera:2018jxt,Ferlito:2017xdq,Bourget:2019aer,Bhardwaj:2019jtr,Bourget:2020gzi,vanBeest:2020civ,Closset:bu,Apruzzi:2019vpe,Apruzzi:2019opn,Apruzzi:2019enx,Closset:2020scj,Closset:2020afy,Collinucci:2021wty,Collinucci:2021ofd}, special attention was given to five-dimensional superconformal
field theories (SCFTs).  In dimension five, the Yang-Mills coupling $g_{\text{YM}}$ is
dimensionful and $\frac{1}{g_{\text{YM}}^2}$ is seen, from the point of view of the superconformal
field theory, as a (instantonic) mass term. This fact has two immediate
consequences. The first one is that the superconformal fixed point can be reached only as
an infinite coupling limit. Secondly, some quantities, such as the
Higgs Branch (HB) of the theory, are protected against RG-flow just until we
keep the coupling finite: at infinite coupling fixed points some
instantonic particles can become massless, opening up new HB directions.

As a consequence, the gauge theory analysis can not be trusted to
investigate the superconformal fixed point, and we have to resort to
different methods to study it. In this sense, great help can come from
string theory: stringy constructions, involving five-branes webs
in type IIB \cite{Aharony:1997bh,Cabrera:2018jxt,Bourget:2019aer,Bourget:2020gzi} or geometric engineering of M-theory on
threefold Calabi-Yau (CY) singularities \cite{Intriligator:1997pq,
Zotto:bj,Apruzzi:2019opn,Apruzzi:2019enx,Closset:bu,Closset:2020scj,Closset:2020afy,Collinucci:2021wty,Collinucci:2021ofd} can be used to study and classify
five-dimensional superconformal fixed points. This latter M-theoretic
geometric approach is well understood for toric CY \cite{Closset:bu}, but
still less studied for other classes of CY (see, e.g., \cite{Zotto:bj,Apruzzi:2019opn,Apruzzi:2019enx} for CY
that are elliptically fibered, and \cite{Closset:2020scj,Closset:2020afy} for the case of Isolated
Hypersurface Singularities (IHS) for recently proposed approaches).

In this paper we are going to focus on the last approach, applying the techniques proposed
in \cite{Collinucci:2021wty,Collinucci:2021ofd}: we will engineer
five-dimensional SCFTs as M-theory on a particular class of IHS, namely
those that can be described as fibrations of deformed ADE
singularities over a complex parameter $w \in \mathbb C_w$. These
threefold CY singularities admit either no crepant\footnote{A crepant
  resolution is a resolution that preserves the triviality of the canonical
bundle of the threefold (namely, the CY condition).} resolution (namely, they are
``terminal'' singularities), or a small crepant resolution (a crepant
resolution whose exceptional locus is in complex codimension two). This
means that we have, from the point of view of the moduli space of vacua of the SCFT, an
empty Coulomb Branch (CB). If the singularity admits a small crepant
resolution, we can still have a non-empty Extended Coulomb Branch (ECB),
parametrized by the K\"ahler volumes of the compact holomorphic curves
(possibly) inflated in the resolution. These K\"ahler parameters will play the role of five-dimensional real mass parameters. The approach of \cite{Collinucci:2021ofd} uses
a type IIA limit of M-theory (involving D6 branes and O$6^{-}$ planes) to rephrase the M-theory dynamics in terms of
 a $\mathcal N = 1, D=7$ supersymmetric gauge theory describing the
D6 branes (plus, possibly, O$6^{-}$ planes) physics. In this picture, the M-theory dynamics is
captured by a complex-valued Higgs field $\Phi \equiv \Phi_1+ i\Phi_2$,
constructed combining two out of the three adjoint scalars
$\Phi_{1},\Phi_{2},\Phi_{3}$ whose eigenvalues describe the D6 branes
position in the transverse directions, as depicted in this table:
\begin{table}[H]
    \centering
    \begin{tabular}{c|c|c|c|c}
    & &$ \mathbb{C}_z$&  & \\
         &  $\mathbb R^{6,1}$ & $\overbrace{\Phi_1 | \Phi_2}$ &$\Phi_3 $\\
         \hline
    D6     &  $\times$ & & 
    \end{tabular}
    \caption{IIA setup dual to M-th. on $\mathbb R^{6,1}\times \text{ADE}$}
    \label{tab:my_label}
\end{table}
From a threefold perspective, we are going to interpret $\Phi$ as a complex deformation and $\Phi_3$ as the small blow-up modes. In particular, the D-term relation:
\begin{equation}
    \comm{\Phi}{\Phi_3}= 0
\end{equation}
shows that $\Phi_3$ picks a constant VEV along the Cartan generators dual to the roots that are being resolved, as explained in deeper detail in \cite{Collinucci:2021ofd}.

We will study and completely determine (as a complex algebraic variety) the Higgs-Branches of two
infinite classes of ADE-fibered IHS singularities\footnote{We follow the notation of \cite{Cecotti:2011rv}.}: the $(A_j,A_l)$ series
\begin{equation}
  \label{AjAk family 1}
  x^2 + y^2 + z^{j+1} + w^{l+1} = 0, \quad (x,y,w,z) \in \mathbb C^4,
\end{equation}
and the $(A_k,D_n)$ series:
\begin{equation}
  \label{AmDn family 1}
  x^2 + zy^{2} + z^{n-1} + w^{k+1}= 0, \quad (x,y,w,z) \in \mathbb C^4. 
\end{equation}
Both the $(A_j,A_l)$ family and the $(A_k,D_n)$ are extensively studied in
type IIB setup, where they realize the corresponding Argyres-Douglas four-dimensional SCFTs, and (apart from the
trivial $(A_1,A_1)$ conifold case) are non-toric, and then difficult to
approach from the M-theory perspective. As the Coulomb Branches of the 5d theories are empty, we interpret the localized modes as hypermultiplets: indeed, anytime we have a resolvable threefold the exceptional curves resolving the singularity are rigid, and consequently the M2 branes states wrapping the curves have to be interpreted as 5d hypermultiplets \cite{Witten}.\\
\indent We stress that the range of application of our method is not limited to the analysis of the $(A_j,A_l)$ and $(A_k,D_n)$ singularities, but can be employed to study \textit{any} one-parameter family of deformed $A_l$ or $D_n$ singularities.\\

The paper is structured as follows. In section \ref{sec:a_j-a_k-series} we focus on the
$(A_j,A_l)$ series. We will quickly recap the construction of \cite{Collinucci:2021wty,Collinucci:2021ofd}, produce an explicit example, and show a simple rule to succinctly analyze all possible cases. We will finally outline an interpretation for the studied HBs in terms of the geometry of nilpotents orbits of simple Lie algebras. 

In section \ref{section: akdncase}, we will focus on the $(A_k,D_n)$ series, recalling how to include
O6$^{-}$ planes in the type IIA limit \cite{Collinucci:2021ofd}, furnishing explicit examples along the way.

We will conclude, in section \ref{T-branes section}, highlighting an interesting T-brane
hierarchy structure, explicit from the point of view of the branes
dynamics, for both the $(A_j,A_l)$ and $(A_k,D_n)$ cases. In section \ref{section: conclusions}, we
will sum up our conclusions and propose possible follow-ups of our work.

We will give the tables for the $(A_k,D_n)$ series and elucidate some technical aspects of our construction in the Appendices.
\section{$\boldsymbol{(A_j,A_l)}$ singularities}
\label{sec:a_j-a_k-series}
In this section we want to completely characterize the Higgs Branches of the 5d SCFTs engineered by M-theory on $(A_j,A_l)$ singularities.\\
\indent In this case, the type IIA limit contains just D6 branes, and no
O6$^{-}$ planes, and hence the analysis is simplified. To take the type IIA limit explicitly, it is
convenient to rewrite \eqref{AjAk family 1} as:
\begin{equation}
  \label{AjAk family 2}
  uv = z^{j+1} + w^{l+1}, \quad u \equiv (x +iy), \quad v \equiv -x+ iy.
\end{equation}
The $\mathbb C^*$ action describing the $\mathbb C^*$ fibration is
\begin{equation}
  \label{Cstar action AjAk}
  u \to \lambda u, \quad v \to \frac{v}{\lambda}, \quad \lambda \in \mathbb C^*,
\end{equation}
and the combination $uv$ appearing on the l.h.s. of \eqref{Cstar action AjAk} is the associated moment map. The degeneracy locus of the
$\mathbb C^*$ fibration corresponds to the zeros of the moment
map, and in the type IIA limit it corresponds to the position of the D6 branes. Consequently, using again \eqref{Cstar action AjAk} the brane locus  ``$\Delta$'' is
\begin{equation}
  \label{Brane locus AjAk}
  \Delta = z^{j+1}+ w^{l+1}  = 0.
\end{equation}

We will now start showing how our techniques work in an explicit example, and then we will generalize to the generic $(A_j,A_l)$ singularity.

\subsection{Warm up: $(A_{14},A_8)$ singularity}
Let us show an example: the singularity $(A_{14},A_8)$ 
\begin{equation}
  \label{Example: A14A8}
  uv = z^{15} + w^9, \quad \quad (u,v,w,z) \in \mathbb C^4. 
\end{equation}
The brane locus is, in this case
\begin{equation}
  \label{Brane locus A14A8}
  \Delta \equiv z^{15}+ w^9 = (z^5+ w^3)(z^5+ e^{\frac{2 \pi i}{3}}w^3)(z^5+ e^{\frac{4 \pi i}{3}}w^3),
  \quad  (w,z) \in \mathbb C^2.
\end{equation}
We want to obtain $\Delta$ as the characteristic polynomial of a ``Higgs field'' $\Phi_{(A_{14},A_8)}$ in the seven-dimensional gauge algebra $\mathfrak g = A_{14}$. More precisely, $\Phi_{(A_{14},A_8)}$ will be a $15
\times 15$ traceless matrix, with matrix entries being polynomials in $w$ \textit{of
  degree at most one}. 
  
  To find such Higgs field, we concentrate on the three irreducible factors of $\Delta$ appearing in  \eqref{Brane locus A14A8}. Each of the irreducible factors of $\Delta$
will correspond to a
diagonal sub-block of the Higgs field, taking value in $\text{Mat}(5,\mathbb
C[w])$. Let's concentrate on the first factor, $z^{5}+w^3$. We have (up to
adjoint action of the subgroup $SU_{\mathbb C}(5) \cong \text{Sl}(5,\mathbb C)$ of the seven-dimensional gauge group $SU_{\mathbb C}(15)$) two possible Higgs field blocks\footnote{Here we are using the fact that we are taking the entries of $\Phi_{(A_{14},A_8)}$ to be polynomial of $w$ of degree at most one.} with characteristic polynomial $(z^5+w^3)$:
\begin{equation}
  \label{Higgs vs T-branes A14A8}
 \mathcal A^{(I)} = \left(
\begin{array}{ccccc}
 0 & w & 0 & 0 & 0 \\
 0 & 0 & 1 & 0 & 0 \\
 0 & 0 & 0 & w & 0 \\
 0 & 0 & 0 & 0 & 1 \\
 -w & 0 & 0 & 0 & 0 \\
\end{array}
\right), \quad  \mathcal A^{(II)} = \left(
\begin{array}{ccccc}
 0 & w & 0 & 0 & 0 \\
 0 & 0 & w & 0 & 0 \\
 0 & 0 & 0 & 1 & 0 \\
 0 & 0 & 0 & 0 & 1 \\
 -w & 0 & 0 & 0 & 0 \\
\end{array}
\right).
\end{equation}
We start choosing $\mathcal A^{(I)}$ as the diagonal block of $\Phi_{(A_{14},A_8)}$ associated to $z^5 + w^3$; we will comment later on the alternative choice 
$\mathcal{A}^{(II)}$. For now, we just notice that $\mathcal{A}^{(I)}$ and $\mathcal{A}^{(II)}$ are distinguished by their Jordan form at $w =0$, ($[2,2,1]$ for $\mathcal{A}^{(I)}$ and $[3,1,1]$ for $\mathcal A^{(II)}$), or, equivalently, by the corresponding nilpotent orbits under $SU_{\mathbb C}(15)$.

We can  then take the block diagonal sum of the $\mathcal A^{(I)}$ corresponding to each factor in \eqref{Brane locus A14A8} obtaining:
\begin{eqnarray}
  \label{Higgs A14A8}
\Phi_{(A_{14},A_{8})}\equiv  \scalemath{0.9}{\left(
\begin{array}{ccccc|ccccc|ccccc}
 \textcolor{red}{0} & \textcolor{red}{w} & \textcolor{red}{0} & \textcolor{red}{0} & \textcolor{red}{0} & 0 & 0 & 0 & 0 & 0 & 0 &
   0 & 0 & 0 & 0 \\
 \textcolor{red}{0} & \textcolor{red}{0} & \textcolor{red}{1} & \textcolor{red}{0} & \textcolor{red}{0} & 0 & 0 & 0 & 0 & 0 & 0 &
   0 & 0 & 0 & 0 \\
 \textcolor{red}{0} & \textcolor{red}{0} & \textcolor{red}{0} & \textcolor{red}{w} & \textcolor{red}{0} & 0 & 0 & 0 & 0 & 0 & 0 &
   0 & 0 & 0 & 0 \\
\textcolor{red}{0} & \textcolor{red}{0} & \textcolor{red}{0} & \textcolor{red}{0} & \textcolor{red}{1} & 0 & 0 & 0 & 0 & 0 & 0 &
   0 & 0 & 0 & 0 \\
\textcolor{red}{-w} & \textcolor{red}{0} & \textcolor{red}{0} & \textcolor{red}{0} & \textcolor{red}{0} & 0 & 0 & 0 & 0 & 0 & 0 &
                                                                                                                                 0 & 0 & 0 & 0 \\
  \hline
 0 & 0 & 0 & 0 & 0 & \textcolor{blue}{0} & \textcolor{blue}{w} & \textcolor{blue}{0} & \textcolor{blue}{0} & \textcolor{blue}{0} & 0 &
   0 & 0 & 0 & 0 \\
 0 & 0 & 0 & 0 & 0 & \textcolor{blue}{0} & \textcolor{blue}{0} & \textcolor{blue}{1} & \textcolor{blue}{0} & \textcolor{blue}{0} & 0 &
   0 & 0 & 0 & 0 \\
 0 & 0 & 0 & 0 & 0 &  \textcolor{blue}{0} & \textcolor{blue}{0} & \textcolor{blue}{0} & \textcolor{blue}{w} & \textcolor{blue}{0} & 0 &
   0 & 0 & 0 & 0 \\
 0 & 0 & 0 & 0 & 0 & \textcolor{blue}{0} & \textcolor{blue}{0} & \textcolor{blue}{0} & \textcolor{blue}{0} & \textcolor{blue}{1} & 0 &
   0 & 0 & 0 & 0 \\
 0 & 0 & 0 & 0 & 0 &  \textcolor{blue}{-e^{\frac{2 \pi i}{3}} w} & \textcolor{blue}{0} &
                                                                        \textcolor{blue}{0} & \textcolor{blue}{0} & \textcolor{blue}{0} & 0 & 0 & 0 & 0 & 0 \\
  \hline
 0 & 0 & 0 & 0 & 0 & 0 & 0 & 0 & 0 & 0 & \textcolor{ForestGreen}{0} & \textcolor{ForestGreen}{w} & \textcolor{ForestGreen}{0} & \textcolor{ForestGreen}{0} & \textcolor{ForestGreen}{0}  \\
 0 & 0 & 0 & 0 & 0 & 0 & 0 & 0 & 0 & 0 &  \textcolor{ForestGreen}{0} & \textcolor{ForestGreen}{0} & \textcolor{ForestGreen}{1} & \textcolor{ForestGreen}{0} & \textcolor{ForestGreen}{0}  \\
 0 & 0 & 0 & 0 & 0 & 0 & 0 & 0 & 0 & 0 & \textcolor{ForestGreen}{0} & \textcolor{ForestGreen}{0} & \textcolor{ForestGreen}{0} & \textcolor{ForestGreen}{w} & \textcolor{ForestGreen}{0}  \\
 0 & 0 & 0 & 0 & 0 & 0 & 0 & 0 & 0 & 0 &  \textcolor{ForestGreen}{0} & \textcolor{ForestGreen}{0} & \textcolor{ForestGreen}{0} & \textcolor{ForestGreen}{0} & \textcolor{ForestGreen}{1}  \\
 0 & 0 & 0 & 0 & 0 & 0 & 0 & 0 & 0 & 0 &  \textcolor{ForestGreen}{-e^{\frac{4 \pi i}{3}}w} & \textcolor{ForestGreen}{0} & \textcolor{ForestGreen}{0} & \textcolor{ForestGreen}{0} & \textcolor{ForestGreen}{0}  \\
\end{array}
\right)},\nonumber \\
\end{eqnarray}
where we multiplied the lower-left entry of each block by an appropriate third root of unity, in such a way to reproduce the factors of \eqref{Brane locus A14A8}.
One can explicitly check that the characteristic polynomial of $\Phi_{(A_{14},A_8)}$ is $\Delta$.

We now show how to count five-dimensional modes. First, we decompose the
oscillations ``$\varphi$'' of the Higgs field and the
seven-dimensional gauge algebra element $g$ according to the block
decomposition of $\Phi_{(A_{14},A_8)}$:
\begin{eqnarray}
  \label{Oscillations block decomposition A14A8}
  \scalemath{0.8}{\Phi_{(A_{14},A_8)}  = \left(
            \begin{array}{ccc}

              \mathcal A_1 & 0 & 0 \\
              0 & \mathcal A_2 & 0 \\
              0 & 0 & \mathcal A_3 \\
            \end{array}
  \right), \quad \varphi = \left(
                                \begin{array}{ccc}
                                  \varphi_{(1,1)} & \varphi_{(1,2)}& \varphi_{(1,3)} \\
                                  \varphi_{(2,1)} & \varphi_{(2,2)} & \varphi_{(2,3)} \\
                                  \varphi_{(3,1)} & \varphi_{(3,2)} & \varphi_{(3,3)} \\
                                \end{array}
  \right), \quad g =  \left(
                              \begin{array}{ccc}
                                \mathcal{G}_{(1,1)} & \mathcal{G}_{(1,2)}&\mathcal{G}_{(1,3)} \\
                                \mathcal{G}_{(2,1)} & \mathcal{G}_{(2,2)} & \mathcal{G}_{(2,3)} \\
                                \mathcal{G}_{(3,1)} & \mathcal{G}_{(3,2)} & \mathcal{G}_{(3,3)} \\
                              \end{array}
  \right), }\nonumber \\
\end{eqnarray}
where we recall that:
\begin{eqnarray}
  \label{Blocks reminder A14A8}
  \scalemath{0.8}{\mathcal A_1 \equiv \left(
\begin{array}{ccccc}
 0 & w & 0 & 0 & 0 \\
 0 & 0 & 1 & 0 & 0 \\
 0 & 0 & 0 & w & 0 \\
 0 & 0 & 0 & 0 & 1 \\
 -w & 0 & 0 & 0 & 0 \\
\end{array}
\right), \quad \mathcal A_2
  \equiv \left(
\begin{array}{ccccc}
 0 & w & 0 & 0 & 0 \\
 0 & 0 & 1 & 0 & 0 \\
 0 & 0 & 0 & w & 0 \\
 0 & 0 & 0 & 0 & 1 \\
 -e^{2 \pi i/3}w & 0 & 0 & 0 & 0 \\
\end{array}
  \right),\nonumber \quad
  \mathcal A_3 \equiv \left(
\begin{array}{ccccc}
 0 & w & 0 & 0 & 0 \\
 0 & 0 & 1 & 0 & 0 \\
 0 & 0 & 0 & w & 0 \\
 0 & 0 & 0 & 0 & 1 \\
 -e^{4 \pi i/3}w & 0 & 0 & 0 & 0 \\
\end{array}
\right)}
\end{eqnarray}
and $\mathcal{G}_{(i,j)} \in \text{Mat}(5,\mathbb C[w])$, $\varphi_{(i,j)} \in \text{Mat}(5,\mathbb C[w])$.
We now have to count deformations $\varphi$ that are not gauge equivalent to zero, modding by linearized gauge transformations
$\comm{\Phi_{(A_{14},A_{8}))}}{g} = \text{ad}_{\Phi_{(A_{14},A_{8})}}(g)$:
\begin{equation}
  \label{Linearized gauge transformation}
  \varphi \sim \varphi + \comm{\Phi_{(A_{14},A_8)}}{g}.
\end{equation}
The block decomposition of $g$ is respected by the action of
$\text{ad}_{\Phi_{(A_{14},A_{8})}}$:
\begin{equation}
  \label{Linearized gauge transformation explicit}
  \comm{\Phi_{(A_{14},A_8)}}{g}= \left(
                              \begin{array}{ccc}
                                \comm{\mathcal A_1}{\mathcal{G}_{(1,1)}} & \mathcal
                                                                 A_1\mathcal{G}_{(1,2)}-
                              \mathcal{G}_{(1,2)} \mathcal A_2&  \mathcal
                                                                 A_1\mathcal{G}_{(1,3)}-
                              \mathcal{G}_{(1,3)} \mathcal A_3\\
                                 \mathcal
                                                                 A_2\mathcal{G}_{(2,1)}-
                              \mathcal{G}_{(2,1)} \mathcal A_1& \comm{\mathcal A_2}{\mathcal{G}_{(2,2)}} &  \mathcal
                                                                 A_2\mathcal{G}_{(2,3)}-
                              \mathcal{G}_{(2,3)} \mathcal A_3 \\
                                 \mathcal
                                                                 A_3\mathcal{G}_{(3,1)}-
                              \mathcal{G}_{(3,1)} \mathcal A_1& \mathcal
                                                                 A_3\mathcal{G}_{(3,2)}-
                              \mathcal{G}_{(3,2)} \mathcal A_2& \comm{\mathcal A_3}{\mathcal{G}_{(3,3)}} \\
                              \end{array}
  \right),
\end{equation}
and we can proceed to gauge fix the oscillations $\varphi$ blockwise. Let's concentrate, for example, on the block $\varphi_{(1,2)}$:
\begin{eqnarray}
  \label{Oscillations block A14A8}
 && \varphi_{(1,2)} \equiv \left(
\begin{array}{ccccc}
 \delta _{1,1} & \delta _{1,2} & \delta
   _{1,3} & \delta _{1,4} & \delta _{1,5} \\
 \delta _{2,1} & \delta _{2,2} & \delta
   _{2,3} & \delta _{2,4} & \delta _{2,5} \\
 \delta _{3,1} & \delta _{3,2} & \delta
   _{3,3} & \delta _{3,4} & \delta _{3,5} \\
 \delta _{4,1} & \delta _{4,2} & \delta
   _{4,3} & \delta _{4,4} & \delta _{4,5} \\
 \delta _{5,1} & \delta _{5,2} & \delta
   _{5,3} & \delta _{5,4} & \delta _{5,5} \\
\end{array}
  \right). \nonumber \\
\end{eqnarray}
The corresponding block of $\comm{\Phi_{(A_{14},A_8)}}{g}$ is:
\begin{equation}
\label{1 Linearized gauge tr. block A14A8}
\mathcal{A}_1 \mathcal{G}_{(1,2)}-
   \mathcal{G}_{(1,2)} \mathcal A_2= A + w B,\nonumber \\
\end{equation}
where we indicated with $A, B$ the following matrices:
\begin{eqnarray}
\label{2 Linearized gauge tr. block A14A8}
 \scalemath{0.8}{A  \equiv  
\left(
\begin{array}{ccccc}
 0 & 0 & -g_{1,7} & 0 & -g_{1,9} \\
 g_{3,6} & g_{3,7} & g_{3,8}-g_{2,7} &
   g_{3,9} & g_{3,10}-g_{2,9} \\
 0 & 0 & -g_{3,7} & 0 & -g_{3,9} \\
 g_{5,6} & g_{5,7} & g_{5,8}-g_{4,7} &
   g_{5,9} & g_{5,10}-g_{4,9} \\
 0 & 0 & -g_{5,7} & 0 & -g_{5,9} \\
\end{array}
\right), 
 B \equiv  \left(
\begin{array}{ccccc}
 (-1)^{2/3} g_{1,10}+g_{2,6} & g_{2,7}-g_{1,6} & g_{2,8} & g_{2,9}-g_{1,8} & g_{2,10} \\
 (-1)^{2/3} g_{2,10} & -g_{2,6} & 0 & -g_{2,8} & 0 \\
 (-1)^{2/3} g_{3,10}+g_{4,6} & g_{4,7}-g_{3,6} & g_{4,8} & g_{4,9}-g_{3,8} & g_{4,10} \\
 (-1)^{2/3} g_{4,10} & -g_{4,6} & 0 & -g_{4,8} & 0 \\
 (-1)^{2/3} g_{5,10}-g_{1,6} & -g_{1,7}-g_{5,6} & -g_{1,8} & -g_{1,9}-g_{5,8} & -g_{1,10} \\
\end{array}
\right)}. \nonumber \\
\end{eqnarray}
Let's start the gauge fixing from the entry $\delta_{1,1}$. The corresponding entry of the linearized gauge transformations is
\begin{equation}
  \label{1 A14A8 linear mode gfixing}
  w((-1)^{2/3} g_{1,10}+g_{2,6}) \in (w) \subset \mathbb C[w],
\end{equation}
and we can, for $w \neq 0$, always put
\begin{equation}
  \label{2 A14A8 linear mode gfixing}
  (-1)^{2/3} g_{1,10}+g_{2,6} = -\frac{\delta_{(1,1)}}{w}, 
\end{equation}
and set to zero the mode in this matrix entry. On the other hand, we can not
gauge fix to zero $\delta_{1,1}$ at the point $w= 0$. From a mathematical
point of view we have, after the gauge fixing:
\begin{equation}
  \label{3 A14A8 linear mode gfixing}
  \delta_{1,1}\in \mathbb C[w] \quad \stackrel{gauge fix}{\longrightarrow} \quad  \delta_{1,1} \in \mathbb C[w]/(w), 
\end{equation}
This means, in the algebraic geometry language, that the mode
$\delta_{1,1}$ is a ``delta function'' localized
at the intersection $w=0$ between the stacks associated to $\mathcal
A_1,\mathcal A_2$ of the D6 branes
system\footnote{All the factors of $\Delta$ intersect (with multiplicity)
  just at $ w=0$.}.

Let's proceed with another gauge fixing. The entry $\delta_{1,3}$ can be completely fixed to zero
choosing
\begin{equation}
  \label{A14A8 complete gfixing}
  g_{1,7}= w g_{2,8}+\delta_{1,3} 
\end{equation}
and does not correspond to any five-dimensional mode. Finally we remark that, proceeding with the gauge-fixing procedure, it might be that a mode $\delta_{i,j}$ cannot be localized at all, the corresponding entry of $\comm{\Phi_{(A_{14},A_8)}}{g}$ being empty. In this case, $\delta_{i,j}$ is a physical mode that cannot be gauge fixed to zero, but that is not localized in five dimensions, and then decouples. We can proceed in a similar way to gauge fix all the modes, obtaining:
\begin{equation}
  \label{Modes position A14A8}
\varphi_{(1,2)}^{\text{gauge fixed}} =  \left(
\begin{array}{ccccc}
 \delta _{1,1} & \delta _{1,2} & 0 & \delta _{1,4} & 0 \\
 0 & 0 & 0 & 0 & 0 \\
 \delta _{3,1} & \delta _{3,2} & \delta _{3,3} & \delta _{3,4} & \delta _{3,5} \\
 0 & 0 & 0 & 0 & 0 \\
 \delta _{5,1} & \gamma  & \delta _{5,3} & \tau  & \delta _{5,5} \\
\end{array}
\right),
\end{equation}
with $\delta_{i,j} \in \mathbb C[w]/(w)$ $\forall i,j$ and $\gamma, \tau
\in \mathbb C[w]/(w^2)$, that hence give two complex dimensional modes each. In total, we have fifteen five-dimensional modes
localized in the block corresponding to $\varphi_{(1,2)}$. The analysis proceeds similarly for the other
blocks.

To completely characterize the Higgs Branch, we now have to understand the
action of the flavor and discrete symmetries on the five-dimensional
modes. We have, if we pick the seven-dimensional gauge group to be
$SU(15)$,
\begin{equation}
  \label{Stabilizers A14A8}
  \text{Stab}(\Phi) \cong \left( U(1)^3/U(1)_{diag}\right) \times \mathbb Z_5,
\end{equation}
where the stabilizers $\text{Stab}(\Phi)$ are computed w.r.t.\ the adjoint
action of the seven-dimensional gauge group $G_{7d}$,
\begin{equation}
  \label{Stabilizers wrt adjoint action}
  \Phi \quad \longrightarrow \quad G \Phi G^{-1}, \quad \quad G \in G_{7d}.
\end{equation}
We identify, following \cite{Collinucci:2021ofd}, $U(1)^3/U(1)_{diag}$ to be the five-dimensional flavor group, and $\mathbb
Z_5$ the five-dimensional discrete gauging group. The flavour group is:
\begin{equation}
  \label{Flavor group A14A8}
G_{\text{flavor}} \equiv  \left(
\begin{array}{ccc}
 e^{i \alpha}\mathbbm{1}_5 &  &  \\
  & e^{i\beta}\mathbbm{1}_5 &  \\
  &  & e^{i\gamma}\mathbbm{1}_5 \\
\end{array}
\right) \in U(1)^3/U(1)_{diag}, \quad \alpha + \beta + \gamma = \frac{2 \pi n}{5},
\end{equation}
while the generator of the discrete gauging is:
\begin{equation}
  \label{Discrete gauging A14A8}
  G_{\text{gauge}}\equiv
  \left(\begin{array}{ccc}
    e^{ \frac{2 \pi i}{5}} \mathbbm{1}_5 &  &  \\
     & \mathbbm{1}_5 &  \\
     &  & \mathbbm{1}_5 \\
  \end{array}
\right) \in \mathbb Z_5.
\end{equation}
Both the discrete gauging group and the flavour group act on the modes by
adjoint action:
\begin{equation}
  \label{Flavor and discrete g. groups action}
  \varphi \to G_{\text{gauge}}\varphi (G_{\text{gauge}})^{-1}, \quad \varphi
  \to G_{\text{flavor}}\varphi (G_{\text{flavor}})^{-1}.
\end{equation}
Then, simply doing a linear algebra computation we have that, e.g., all the modes in $\varphi_{(1,2)}$ transform linearly with charge one
under $U(1)_{\alpha},$ they transform with charge minus one under $U(1)_{\beta}$, are
singlets w.r.t. $U(1)_{\gamma}$, and transform linearly under $\mathbb Z_5$, each one being multiplied by a fifth root of unity. The
following table, where we filled each block with the number of localized
five-dimensional modes, permits (together with \eqref{Flavor group A14A8} and \eqref{Discrete gauging A14A8}) to extract  the action of the flavor and
discrete group on the modes, and then to characterize the Higgs branch as a complex
variety:
\begin{equation}
 \label{Modes A14A8}
 \text{modes }= \left(
    \begin{array}{c|c|c}
      8 & 15 & 15  \\
      \hline
      15 & 8 & 15  \\
      \hline
      15 & 15 & 8  \\
    \end{array}
  \right).
\end{equation}
\begin{table}[H]
  \centering
  \begin{tabular}{ |p{2.cm}||p{1.5cm}|p{1.5cm}|p{1.5cm}|p{1.5cm}|p{1.5cm}|}
    \hline
    \textbf{Block} & $\boldsymbol{n_{modes}}$ & $\boldsymbol{U(1)_{\alpha}}$& $\boldsymbol{U(1)_{\beta}}$  & $\boldsymbol{U(1)_{\gamma}}$ & $\boldsymbol{\mathbb{Z}_5}$\\
    \hline
    $\varphi_{(1,1)}$ & 8     &  0      & 0        & 0        &     -         \\
    $\varphi_{(2,2)}$ & 8     &  0      & 0        & 0        &     -         \\
    $\varphi_{(3,3)}$ & 8     &  0      & 0        & 0        &     -         \\   
    \hline
    $\varphi_{(1,2)}$ & 15     &  1      & -1        & 0        &     \checkmark\\
    $\varphi_{(1,3)}$ & 15     &  1      & 0        & -1        &     \checkmark\\
    $\varphi_{(2,3)}$ & 15     &  0      & 1        & -1        &     -\\
    \hline                                                             
    $\varphi_{(2,1)}$ & 15     &  -1      & 1        & 0        &     \checkmark\\
    $\varphi_{(3,1)}$ & 15     &  -1      & 0        & 1        &     \checkmark\\
    $\varphi_{(3,2)}$ & 15     &  0      & -1        & 1        &     -\\  
    \hline
  \end{tabular}
\end{table}
Summing up, we have a Higgs branch of quaternionic dimension 57 (as it can
be computed also with alternative methods \cite{Xie}), and the Higgs
branch is
\begin{equation}
  \label{HB A14A8}
  \text{HB} = \mathbb C^{54} \times \frac{\mathbb C^{60}}{\mathbb
  Z_5},
\end{equation}
with $\mathbb Z_5$ multiplying by $e^{\frac{2 \pi i}{5}}$ the first half of the coordinates of $\mathbb
C^{60}$, and by $e^{-\frac{2 \pi i}{5}}$ the second half.

\subsection{General case: $(A_j,A_l)$ singularity}
In this section we aim at obtaining the Higgs Branch of M-theory on a generic $(A_j,A_l)$ singularity, given by the equation \eqref{AjAk family 2}, that we repeat here for convenience:
\begin{equation*}
  uv = z^{j+1} + w^{l+1}.
\end{equation*}
In order to achieve this task, we decompose (as in \eqref{Brane locus A14A8}) the brane
locus in irreducible factors (namely, polynomials in $(w,z)$ that do not
admit further factorization). Each factor corresponds, geometrically, to an
irreducible component of the brane locus (seen as a one-dimensional subvariety
of $\mathbb C_{w}\times \mathbb C_{z}$).

We write $(A_j,A_l)$ as
$(A_{mp-1},A_{mq-1})$, with $p,q$ coprimes, $p \geq q$, and $m =
\gcd(j+1,l+1)$. It then becomes manifest that we can always factor the
brane locus as follows:
\begin{equation}
  \label{AjAk family 3}
  \Delta = z^{m p} + w^{m q} = \prod_{s=1}^m(z^p + e^{2 \pi i s/m}w^q).
\end{equation}
The factor $(z^p + e^{2 \pi i s/m}w^q)$ in \eqref{AjAk family 3} can
be realised, for all the $(p,q)$, as the characteristic polynomial of a $p
\times p$ matrix $\mathcal A_s$, with matrix entries being polynomials in $w$ \textit{of
  degree at most one.}
  
  The blocks
``$\mathcal A_s$'' (they are, indeed, characterized by the four integers $p,q,s,m$, appearing in each
factor of \eqref{AjAk family 3} but we omit $p,q,m$ for ease of notation) whose characteristic polynomials are the irreducible factors appearing in \eqref{AjAk family 3}
can be put in the following canonical shape\footnote{Notice that the canonical form \eqref{Higgs block Ak} is precisely in the shape of a companion matrix, as usually understood in the mathematical literature.}:

\begin{equation}
  \label{Higgs block Ak}
  \mathcal A_s(w) =\left(
    \begin{array}{ccccc}
      0 & \ast & 0 & \cdots & 0 \\
      0 & 0 & \ast  & 0 & 0\\
      \vdots & 0 & \ddots & \ddots & 0  \\
      0 & 0 & 0 & 0 & \ast \\
      -e^{2 \pi i s/m}w & 0 & 0 & 0& 0 \\
    \end{array} \right),
\end{equation}
where the $\ast$ entries are filled either with $w,$
or are constants (that can be set to 1); to reproduce the right
characteristic polynomial, we have to fill $q-1$
$\ast$-entries with 
``$w$''.\\
\indent Depending on the position where we place the ``$w$'', one has a different number of zero-modes. D-brane states realizing the same brane locus (or, analogously, dual to the same M-theory geometry) but with a different number of zero-modes are known in this context as \textit{T-brane states}.

We found a nice criterium to
understand (without performing any computation\footnote{As we will explain
  below, one can always run the gauge fixing analysis to check this a posteriori. The
  codimension formula for $A_j$-fibered threefolds simply shorten this
  process. Unfortunately, the same argument can not be applied in the $D_k$
case.}) if the chosen $\mathcal A_s$ describes a T-brane background. The argument holds more generally for any Higgs field $\Phi$. Indeed, we can think of $\mathcal A_s$ itself as the Higgs field associated to the singularity $uv = \chi(\mathcal A_s)$, with $\mathcal A_s \in A_{p-1}$ and $\chi$ the characteristic polynomial. Keeping the
characteristic polynomial fixed, we associate to $\Phi$ the nilpotent orbit $\mathcal{O}_{0} $ obtained acting with the seven-dimensional gauge group on
$\Phi\rvert_{w=0}$. We found  that the number $n_{\text{ind}}$ of linearly independent elements of the seven-dimensional gauge algebra ``$\mathfrak{g}$'' supporting a five-dimensional zero-mode always equals the complex codimension of
$\mathcal{O}_{0}$ in the nilpotent cone of $\mathfrak{g}$:
\begin{equation}
\label{codimension formula}
    n_{\text{ind}} = \text{cod}_{\mathbb C}\Big(\mathcal{O}_{0}  \hookrightarrow \mathfrak{g}\Big).
\end{equation}
\eqref{codimension formula} holds also for the Higgs fields associated to the threefolds analyzed in Section \ref{section: akdncase}, namely one-parameter families of $D_n$ singularities. Notice that if there are 5d localized modes supported on $\mathbb{C}[w]/(w^k)$ with $k>1$, then $n_{\text{ind}}$ \textit{does not} coincide with the total number of 5d modes localized in the Higgs background.

\indent Furthermore, we found that in all the $(A_j,A_l)$ cases the Higgs maximizing $n_{\text{ind}}$ also
maximizes the total number of five-dimensional modes. Consequently, the Higgs field $\Phi$ displaying the maximum number of five-dimensional modes is the block-diagonal sum of blocks $\mathcal A_s$  corresponding to
the $\mathcal{O}_{0}$ that sits at
the lowest position in the $A_{p-1}$ nilpotent orbits Hasse diagram, while
the other Higgs fields have obstructed five-dimensional modes\footnote{
  A more precise geometric proof of this is given in the Appendix B
}.
A first example is \eqref{Higgs vs T-branes A14A8}, the leftmost block has no obstructed deformation because the corresponding orbit [2,2,1] sits below [3,1,1] in the $A_5$ Hasse diagram. Another example is the one we get considering the case $(m=1,p=5, q =2)$ analyzed in
   \cite{Collinucci:2021ofd}. The threefold is:
   \begin{equation}
       \label{A4A1}
       uv = z^5 + w^2, \quad (u,v,w,z) \in \mathbb C^4.
   \end{equation}
   We have two possible Higgs backgrounds\footnote{With polynomial entries of degree at most one in $w$.} reproducing the same geometry:
\begin{equation}
  \label{Higgs block p=5 q=2}
  \mathcal A^{(I)} =\left(
    \begin{array}{ccccc}
      0 & 1 & 0 & 0 & 0 \\
      0 & 0 & w  & 0 & 0\\
      0 & 0 & 0 & 1 & 0  \\
      0 & 0 & 0 & 0 & 1 \\
      -w & 0 & 0 & 0& 0 \\
    \end{array} \right), \quad   \mathcal A^{(II)} =\left(
    \begin{array}{ccccc}
      0 & w & 0 & 0 & 0 \\
      0 & 0 & 1  & 0 & 0\\
      0 & 0 & 0 & 1 & 0  \\
      0 & 0 & 0 & 0 & 1 \\
      -w & 0 & 0 & 0& 0 \\
    \end{array} \right).
\end{equation}
If we want to have no obstructed five-dimensional modes we have to pick the first Higgs, $\mathcal A^{(I)}$. Indeed, the orbit corresponding
to $\mathcal A^{(I)}$ is, in \cite{Collingwood} notations, the
one with partition
$[3,2]$, that is below the
orbit $[4,1]$ (associated to $\mathcal A^{(II)}$) in the $A_4$ nilpotent orbits Hasse diagram. It also turns
out that $[3,2]$ is the tiniest orbit that contains a matrix (that is, $\mathcal A_s\rvert_{w=
0}$)
realizing $(z^5 + w^2)$ as its characteristic polynomial.

Summing up, \textit{if we want to maximize five-dimensional modes}, for a given factor $z^p + e^{2 \pi i s/m}w^q$ in \eqref{AjAk family 3} we pick the block $\mathcal A_s$ such that
$\text{cod}_{\mathbb C}\mathcal{O}_{0} \hookrightarrow \mathcal
N_{A_{p-1}}$ is maximized (with $\mathcal
N_{A_{p-1}}$ the nilpotent cone of $A_{p-1}$). To obtain the $m$ factors of the brane locus corresponding to the full Higgs field, we
take the block direct sum of all the $\mathcal A_s$ blocks\footnote{The
  integers $p,q$ are the same for all the blocks, the phase $e^{2 \pi i s
    /m}$ multiplying the lowest-left entry in \eqref{Higgs block Ak} is opportunely tuned in such a way that each block $\mathcal A_s$
  reproduces each of the factors of
\eqref{AjAk family 3}.}:
\begin{equation}
\label{Higgs block decomposition AjAk}
  \Phi_{(A_{mp-1},A_{mq-1})}\equiv
\arraycolsep=9pt\def\arraystretch{1.4}
\underbrace{\left(\begin{array}{c|c|c|c}
\textcolor{blue}{\mathcal A_{s=1}}&
                           \mathbb 0_{p} &  \mathbb 0_{p} & \mathbb 0_p\\
                    \hline
                    \mathbb 0_p&\textcolor{red}{\mathcal A_{s=2}}  & \vdots \\
                    \hline
                        & \vdots & \ddots&
                                                                       \mathbb 0_{p}  \\
                    \hline
 & \vdots & \vdots & \textcolor{ForestGreen}{\mathcal A_{s=m}} 
\end{array} \right)}_{m \text{ blocks}}.
\end{equation}
The previous procedure applies similarly for all the $(A_{mp-1},A_{mq-1})$
singularities, and we can describe it in general terms as follows:
\begin{enumerate}
\item the five-dimensional modes localize with the following pattern:
\begin{equation} \label{Final answer AjAk}
\varphi \equiv \arraycolsep=9pt\def\arraystretch{1.4}
\underbrace{\left(\begin{array}{c|cc|c}
\textcolor{blue}{(p-1)(q-1) \text{modes} }& p\cdot q \text{ modes}
 & \cdots &  p\cdot q \text{ modes} \\
 p\cdot q \text{ modes}& \ddots & & \vdots \\
 \vdots &  & \ddots &  p\cdot q \text{ modes} \\
 p\cdot q \text{ modes} & \hdots  & p\cdot q \text{ modes} & \textcolor{ForestGreen}{ (p-1)(q-1) \text{ modes}}\\
\end{array} \right)}_{m \text{ blocks}};
\end{equation}
\item the discrete group is always isomorphic to $\mathbb Z_p$. We can pick
  the generator to be:
\begin{equation}
\label{Discrete group general AjAk}
G_{\text{gauge}}\equiv \arraycolsep=9pt\def\arraystretch{1.4}
\underbrace{\left(\begin{array}{c|c|c|c}
\textcolor{blue}{e^{\frac{2 \pi i}{p}}\mathbbm{1}_p}&
                           \mathbb 0_{p} &  \mathbb 0_{p} & \mathbb 0_p\\
                    \hline
                    \mathbb 0_p&\textcolor{red}{\mathbbm{1}_p} & \vdots \\
                    \hline
                        & \vdots & \ddots&
                                                                       \mathbb 0_{p}  \\
                    \hline
 & \vdots & \vdots & \textcolor{ForestGreen}{\mathbbm{1}_p} 
\end{array} \right)}_{m \text{ blocks}};
\end{equation}
\item the flavor group matrix $G_{\text{flavor}} \in U(1)^m/U(1)_{diag}$ is: 
  \begin{equation}
  \label{Flavor group general AjAk}
G_{\text{flavor}}\equiv \arraycolsep=9pt\def\arraystretch{1.4}
\underbrace{\left(\begin{array}{c|c|c|c}
\textcolor{blue}{e^{i \alpha_1}\mathbbm{1}_p}&
                           \mathbb 0_{p} &  \hdots & \mathbb 0_p\\
                    \hline
                    \mathbb 0_p&\textcolor{red}{e^{i \alpha_2}\mathbbm{1}_p} & \vdots&\vdots \\
                    \hline
 \vdots                       & \vdots & \ddots&
                                                                       \mathbb 0_{p}  \\
                    \hline
 \mathbb 0_p& \hdots & \mathbb 0_p & \textcolor{ForestGreen}{e^{i \alpha_m}\mathbbm{1}_p} 
\end{array} \right)}_{m \text{ blocks}}, \quad
 \sum_{s=1}^m\alpha_s = \frac{2 \pi n}{p}; 
\end{equation}
\item $G_{\text{gauge}},G_{\text{flavor}}$ act on the modes by adjoint action:
  \begin{equation}
    \label{Adjoint action AjAk 2}
    \varphi \to G_{\text{gauge}}\varphi(G_{\text{gauge}})^{-1}, \quad \quad
    \quad \varphi \to G_{\text{flavor}}\varphi(G_{\text{flavor}})^{-1};
  \end{equation}
\end{enumerate}
The data in the matrix \eqref{Final answer AjAk} allows us to completely reconstruct the Higgs Branches \textit{as complex varieties}\footnote{In particular, we cannot determine the hyperk\"ahler metric.}. That matrix and the shape of the flavour and discrete gauging are already sufficient to
reconstruct completely the Higgs Branch, and determine the action of the
flavor group.

For the $(A_{mp-1},A_{mq-1})$, we
can do more, and recollect the result in a closed compact formula.
Looking at \eqref{Final answer AjAk}, we get the following general formula for the Higgs branch:
\begin{equation}
  \label{HB general expression AjAk}
  \text{HB}_{m,p,q} = \mathbb C^{\left(m^2-2 m+2\right) p q+m (1-p-q)}\times \frac{\mathbb
    C^{2(m-1)p q}}{\mathbb Z_p}. 
\end{equation}

The $\mathbb Z_p$ acts multiplying the first half of the $\mathbb C^{2(m-1)p q}$ complex coordinates (corresponding to, say, the chiral
complex scalars inside the hypers) by $e^{\frac{2 \pi i}{p}}$, and the second half of the $
\mathbb C^{2(m-1)p q}$ complex coordinates (that correspond to the complex
scalars in the anti-chiral part of the hyper) by $e^{-\frac{2 \pi i}{p}}$.  
One can readily check that \eqref{HB A14A8} satisfies \eqref{HB general expression AjAk}.

The flavor group is $U(1)^{m}/U(1)_{diag}$, as we saw in \eqref{Flavor group general AjAk}. These abelian factors
correspond, in the resolved CY geometry, to the $\mathbb P^1$s inflated in the
resolution. These $\mathbb P^1$s correspond to the roots at positions $r\times p$, with
$r=1,..,m-1$ in the Dynkin diagram of the $A_{mp-1}$ ADE singularity. For example, in the case of the
$(A_{14},A_{8})$ singularity analyzed in section \eqref{Example: A14A8}, the resolved $\mathbb P^1$s correspond to the node at position five and at position ten of the $A_{14}$ Dynkin diagram:\footnote{For the $A_k$ families case (namely, if we have no O6$^{-}$ planes in the IIA limit), the correspondence between nodes of the Dynkin diagram and $U(1)$ subgroups of $SU(k)$ is particularly easy to read. The node at position $n$ in the Dynkin diagram corresponds to the Cartan element 
\begin{eqnarray*}
    (h_n)_{i,j} = \scalemath{0.8}{\begin{cases}
    1, \quad i = j = n, \nonumber \\
    -1, \quad i = j = n+1, \nonumber \\
    0 \quad \text{otherwise.} \nonumber \\
    \end{cases}
    }
\end{eqnarray*}
In our notation, the exponential of $h_n$ corresponds then to the antidiagonal
  combination $\alpha_{i+1} = -
  \alpha_{i},$ and $\alpha_{j} = 0$ $\forall j \neq i, i+1$ of the flavor group.}
  \begin{figure}[H]
  \begin{center}
      \includegraphics[scale=0.25]{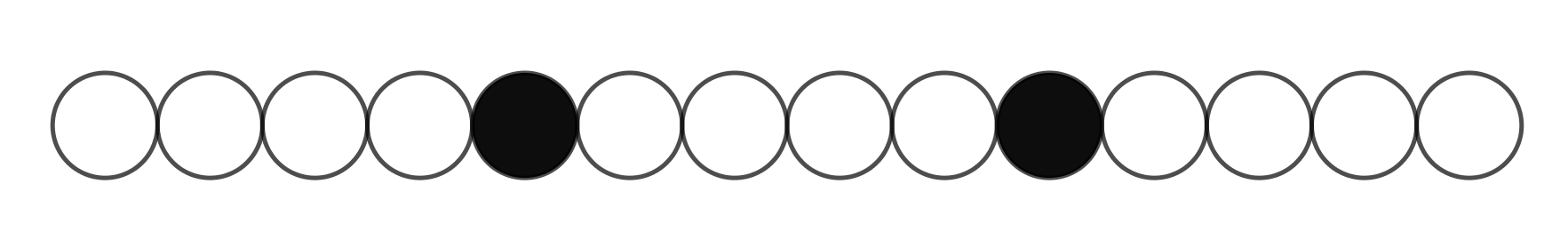}
  \end{center}
  \caption*{The $A_{14}$ Dynkin diagram, with the resolved nodes in black}
  \end{figure}
Each of the flavor $U(1)$ of \eqref{Flavor group general AjAk} acts linearly, with charge one, on
$m (q-1)(p-1)$ modes:
\begin{equation}
  \label{Flavor group action AjAk 2}
  Q_i \to e^{i \alpha_s}Q_{i}, \quad \tilde{Q}_i \to e^{-i
    \alpha_s}\tilde{Q}_{i}, \quad i =1,...,n_{\text{charged hypers}}, \quad s = 1,...,m. 
\end{equation}
We checked explicitly the formula \eqref{HB general expression AjAk} in a
large number of examples.

To conclude, in view of the (more complicated) case of the $(A_k,D_n)$ singularities, we
quickly recap our strategy to get to \eqref{HB general expression AjAk}. Given, as input datum, the equation of the $(A_{mp-1},A_{mq-1})$ singularity:
\begin{enumerate}
\item We \textit{computed the brane locus} $\Delta$ looking where the $\mathbb C^*$ fibers of
  the threefold degenerate.
\item We \textit{factored the brane locus} \eqref{Brane locus AjAk} in polynomials
  that can be represented by the characteristic polynomials of a traceless
  matrix $\mathcal A_s$ with entries being
  $w-$dependent polynomials of degree at most one. We found that \textit{any} polynomial that
  enters in the factorization of the brane locus of the
  $(A_{mp-1},A_{mq-1})$ singularity is the characteristic polynomial of
  some block $\mathcal A_s$ of the shape \eqref{Higgs block Ak}.
  
\item We  \textit{counted the number of five-dimensional modes that are localized in the diagonal blocks}. More precisely, each of the  $\mathcal A_s$ selects a minimal $\mathfrak u(p) \hookrightarrow
  A_{mp-1}$ subalgebra that corresponds to the block containing $\mathcal A_s$ 
  \eqref{Final answer AjAk} (we
  highlighted the $\mathfrak u (p)$ subalgebras corresponding to the various $\mathcal A_s$ with
  different colours in \eqref{Higgs block decomposition AjAk}). The
  localization of modes inside a certain $\mathfrak u(p)$ subalgebra is
  determined  just by the corresponding block $\mathcal A_s$, and is always
  the same for all the $s$.
\item We \textit{counted the number of five-dimensional modes that a pair
  $\textcolor{blue}{\mathcal A_s}$, $\textcolor{red}{\mathcal A_s}$ localizes in the
  corresponding off-diagonal blocks} of the block decomposition of the
  $\mathfrak{sl}(mp,\mathbb C[w])$ matrix (see the equation below):
  \begin{equation} 
  \label{Off-diagoanl block polynomial interpolation AjAk}
\arraycolsep=9pt\def\arraystretch{1.4}
\left(\begin{array}{c|cc|c}
\textcolor{blue}{ (p-1)(q-1) \text{ modes} }& \hdots
 & \cdots & \textcolor{blue}{ p\cdot}\textcolor{red}{ q \text{ modes}} \\
 \vdots & \ddots & & \vdots \\
 \vdots & & \ddots & \vdots \\
 \textcolor{blue}{ p\cdot}\textcolor{red}{ q \text{ modes}}& \hdots & \hdots & \textcolor{red}{ (p-1)(q-1) \text{ modes}}\\
\end{array} \right).
\end{equation}

\end{enumerate}

\section{$\boldsymbol{(A_k,D_n)}$ singularities}
\label{section: akdncase}
In the course of this section, walking along the path undertaken in the preceding pages, we aim at achieving the following results: explicitly construct Higgs field configurations for all the  5d $\mathcal{N}=1$ theories engineered from M-theory on $(A_k,D_n)$ singularities, compute their massless spectrum and determine the continuous and discrete charges, therefore completely characterizing their Higgs Branches.\\

\indent The standard algebraic expression for the $(A_k,D_n)$ singularities reads:
\begin{equation}
    x^2+zy^2+z^{n-1}+w^{k+1}=0, \quad \quad (x,y,w,z) \in \mathbb C^4.
\end{equation}
\indent We should realize $(A_k,D_n)$ singularities as one-parameter families of deformed $D_n$ singularities (with $w$ being the deformation parameter), generalizing the technique already employed in \cite{Collinucci:2021ofd} for flops of length two.
\indent The starting point is a singular threefold, characterized as a deformed $D_n$ family, that constitutes the compactification space of our M-theory setup. As we did in the $(A_j,A_l)$ case, we can make use of the M-theory/Type IIA duality relating a $\mathbb{C}^*$-fibered threefold, namely our deformed $D_n$ family, to a type IIA setup involving D6 branes and O$6^-$ planes, the latter being the new ingredient with respect to the deformed-$A_l$ cases of the previous section.\\
\indent More precisely, to engineer the deformed $D_n$ singularities we consider a stack of $n$ D6/image-D6 pairs, placed on top of a O$6^-$ plane \cite{Collinucci:2008pf} and therefore generating a $SO(2n)$ gauge group in the 7d $\mathcal{N}=1$ theory living on the branes worldvolume. This configuration possesses three real adjoint scalar fields $\Phi_1,\Phi_2,\Phi_3$, belonging to the vector multiplet, describing motion transverse to the brane worldvolume: as in the case of $(A_j,A_l)$ singularities, we package $\Phi_1$ and $\Phi_2$ into a single complex scalar field $\Phi= \Phi_1+i\Phi_2$, depending polynomially on a complex deformation parameter $w$, which can be explicitly described by a $2n\times 2n$ matrix in $\mathfrak{so}(2n)$. The structure of the allowed resolution fixes the scalar field $\Phi_3$, which picks a constant VEV along the Cartan generators dual to the roots that are being resolved. Consequently, the block structure of the Higgs field $\Phi$ describing the deformation of the D6 brane stack is fixed by the relation $\left[\Phi,\Phi_3 \right] = 0$. In this perspective, we pick the following standard basis for the matrices in the $\mathfrak{so}(2n)$ algebra:
\begin{equation}\label{so basis}
\Phi = \left(\begin{array}{cc}
A & B \\
C & -A^t
\end{array} \right), \quad\quad \text{with } B, C \text{ skew-symmetric}.
\end{equation}
As a result, a generic matrix $G$ in the basis \eqref{so basis} belongs to $\mathfrak{so}(2n)$ if and only if it satisfies:
\begin{equation}
G\cdot \mathcal{Q}+\mathcal{Q}\cdot G^t = 0,
\end{equation}
with $\mathcal{Q}$ a quadratic form given by the matrix:
\begin{equation}
\mathcal{Q} = \left(\begin{array}{cc}
0 & \mathbbm{1}_{n\times n} \\
\mathbbm{1}_{n \times n} & 0 \\
\end{array} \right).
\end{equation}
In the following, we will always put the Higgs in the form \eqref{so basis}, or in a block-diagonal form such that every block is in the form \eqref{so basis}.\\
\indent From a physical point of view, switching on a non trivial VEV for $\Phi$ deforms the stack of D6 branes, breaking the $SO(2n)$ group to the subgroup preserved by $\Phi$, and giving rise to a $D_n$ deformed family. The deformation depends on a complex parameter $w$ appearing in the VEV of the Higgs field $\Phi(w)$. The explicit connection between the Higgs VEV describing the dynamics of M-theory on the $(A_k,D_n)$ singularities and the shape of the $D_n$ fibration can be nicely expressed in the following terms\footnote{For threefolds other than the ones that describe M-theory dynamics on $(A_k,D_n)$, there can be additional deformation terms in (\ref{deformed Dn}). For a generic expression in the case of a singularity admitting a single flop we refer to \cite{Collinucci:2021ofd}.}:
\thispagestyle{plain}
\begin{equation}\label{deformed Dn}
x^2+zy^2+\frac{\Delta(z,w)}{z} \quad=\quad x^2+zy^2+\frac{\sqrt{\text{det}(z\mathbbm{1}+\Phi(w)^2)}}{z} \quad=\quad 0,
\end{equation}
where $\Delta(z,w)$ is the locus of D6 branes after a VEV for $\Phi$ has been switched on. Introducing an orientifold-covariant coordinate $\xi$, we can rewrite $z$ as:
\begin{equation}
\label{orientifold coordinates}
z = \xi^2, \quad\quad \text{where }\xi\rightarrow -\xi \text{ under the orientifold projection.}
\end{equation}
Similarly to \eqref{Cstar action AjAk}, the $\mathbb C^*$ fibration basis is $(w,\xi) \in \mathbb C^2$, with fibral coordinates $x,y$ constrained by \eqref{deformed Dn} and the brane locus is 
\begin{equation}
\label{deformed Dn 2} 
\Delta(\xi^2,w) = \text{Disc}_y\Big(\xi^2y^2+\frac{\text{det}(\xi^2\mathbbm{1}+\Phi(w))}{\xi^2}\Big) = \text{det}(\xi^2\mathbbm{1}+\Phi(w)),
\end{equation}
where $\text{Disc}_y$ indicates the discriminant with respect to $y$ (thus justifying the $\Delta$ symbol).\\
\indent At this stage, the story proceeds exactly as in the deformed $A_l$ case: we pick a $(A_k,D_n)$ singularity, and solely by looking at the brane locus we are able to predict which 2-cycles can be resolved and which cannot, thus constraining the block structure of $\Phi$. \\
\indent In order to fully characterize the Higgs Branches of M-theory on all the $(A_k,D_n)$ singularities we should then resort to the recipe outlined at the end of section \ref{sec:a_j-a_k-series}.
All the required steps are straightforward and can be performed exactly following the strategy adopted for the $(A_j,A_l)$ singularities, except the second step, in which a caveat applies: in general, not all possible factorizations into \textit{irreducible polynomials} of the brane locus of a $D_n$ deformed family can be translated into a viable Higgs field. Let us explain this crucial point in more detail. The key ingredient is the relation between the brane locus and the Higgs field $\Phi$:
\begin{equation}\label{brane locus det}
\Delta(\xi^2,w) = \text{det}(\xi^2\mathbbm{1}+\Phi(w)).
\end{equation}
Suppose that the brane locus can be factorized in irreducible holomorphic polynomials of the form:
\begin{equation}
\Delta(\xi^2,w) = P_1(\xi^2,w)\ldots P_m(\xi^2,w).
\end{equation}
Then, according to (\ref{brane locus det}), we would be tempted to build a Higgs field made up of $m$ blocks $\mathcal{B}^{(j)}$, with $j=1,\ldots m$, in some basis of $\mathfrak{so}(2n)$, each contributing a factor $P_j(\xi^2,w)$ to the characteristic polynomial, namely:
\begin{equation}\label{Higgs D4}
\Phi = \left(\begin{array}{c|ccc|c}
\mathcal{B}^{(1)} & 0 & \cdots & & 0 \\
\hline
0 & \ddots & & & \\
\vdots &  & \ddots & & \vdots \\
 &  & & \ddots &  \\
 \hline
0 & & & & \mathcal{B}^{(m)} \\
\end{array} \right), \quad\quad \text{with: } \begin{cases}
\chi(\mathcal{B}^{(1)}) = P_1(\xi^2,w), \\
\quad \vdots \\
\chi(\mathcal{B}^{m)}) = P_m(\xi^2,w), \\
\end{cases}
\end{equation}
where $\chi$ indicates the characteristic polynomial.\\
\indent It turns out that in general this is \textit{not} possible\footnote{This is a known fact in the mathematical literature, as there are no orthogonal companion matrices for orthogonal matrices. This means that, contrarily to the $(A_j,A_l)$ series, a canonical representative for a block in a subalgebra of $\mathfrak{so}(2n)$ with arbitrary characteristic polynomial does not exist.}, meaning that a completely generic irreducible polynomial $P_j(\xi^2,w)$ does not have a counterpart in terms of the characteristic polynomial of a block living in a subalgebra of $\mathfrak{so}(2n)$: as a result, the powerful and general (irreducible polynomial)$\leftrightarrow$(block) correspondence that enabled us to analyze the $(A_j,A_l)$ singularities is broken in the $(A_k,D_n)$ cases. There is, however, some good news: the correspondence is not completely disrupted, and we can reverse the logic of the argument asking the question: is there a way to determine which polynomials in the factorization of the brane locus can be built as the characteristic polynomial of blocks in subalgebras of $\mathfrak{so}(2n)$, and which cannot?\\
\indent We will now show that, for $(A_k,D_n)$, this is indeed possible: we can hence proceed in giving a shortlist of necessary and sufficient blocks needed to reconstruct the brane loci of \textit{all} the $(A_k,D_n)$ singularities.\\
\indent In doing so, we define an \textit{irreducible block} $\mathcal{B}$ as follows:\\

\indent Let $\mathcal{B}$ live in a subalgebra of $\mathfrak{so}(2n)$, and its characteristic polynomial be a polynomial $P(\xi^2,w)$. In general, $P(\xi^2,w)$ might be algebraically reducible and decomposable into factors, but suppose that \textit{at least one} of these factors \textit{cannot} be realized as the characteristic polynomial of a block living in a subalgebra of $\mathfrak{so}(2n)$. Then we say that $\mathcal{B}$ is an \textit{irreducible block}.\\

The list of irreducible blocks also fixes a list of types of polynomials, i.e\ the characteristic polynomial of each block, in which the brane locus $\Delta(\xi^2,w)$ can be consistently factorized. \\
\indent Let us illustrate this concept with a simple example: suppose that in the brane locus factorization of a $(A_k,D_n)$ singularity a factor of the following form appears, corresponding to a block living in a $\mathfrak{so}(4)$ subalgebra of $\mathfrak{so}(2n)$:
\begin{equation}
    P(\xi^2,w) = \xi^4+w^2.
\end{equation}
We would be tempted, on the algebraic level, to go on with the decomposition and write it as the product of two factors, each corresponding to a block in $\mathfrak{so}(2) \subset \mathfrak{so}(2n)$:
\begin{equation}
    P(\xi^2,w) = \xi^4+w^2 = \underbrace{(\xi^2+iw)}_{P_1}\underbrace{(\xi^2-iw)}_{P_2}.
\end{equation}
However, an explicit computation shows that this is \textit{not} possible, i.e.\ there does not exist any holomorphic block in $\mathfrak{so}(2) \subset \mathfrak{so}(2n)$ such that its characteristic polynomial is $P_1$ or $P_2$.\\
\indent Luckily, with some work it is possible to classify the irreducible blocks that are needed to build \textit{all} the Higgs configurations for the $(A_k,D_n)$ singularities that we are interested in\footnote{We have checked this explicitly for $(A_k,D_n)$ singularities with $k$ and $n$ up to the hundreds, and we see no reason not to conjecture that our classification holds for any $k$ and $n$.}.  In Table \ref{Table 1} we list the corresponding characteristic polynomials, which are the polynomials that appear in the factorization of the brane locus, as well as the minimal subalgebras $\mathfrak{s}_{\text{min}}$ of $\mathfrak{so}(2n)$ containing the blocks:
\renewcommand{\arraystretch}{1.5}
\begin{table}[H]
\begin{center}
\begin{tabular}{|c|c|c|}
\hline
$\boldsymbol{\mathcal{B}}$&   $\boldsymbol{\chi(\mathcal{B})}$ & $\boldsymbol{\mathfrak{s}_{\text{min}}}$\\
\hline
(a) & $\left(\xi ^{2 r+1}+c_1 w^t\right) \left(\xi ^{2 r+1}-c_1 w^t\right)$ & $\mathfrak{su}(2r+1)$ \\
\hline
(b) &    $\xi^2 \left(\xi ^{2 r}+c_2 w^{2 t+1}\right)$ 
     & $\mathfrak{so}(2r+2)$ \\
\hline
(c) & \begin{tabular}{c}$\xi^{4 r}+c_3 w^{2t+1}\xi ^{2r}+c_4 w^{2(2t+1)}$  \\
 $(r,2t+1)$ coprime
		\end{tabular} & $\mathfrak{so}(4r)$ \\
\hline
\end{tabular}
\caption{Irreducible blocks and minimal subalgebras.}
\label{Table 1}
\end{center}
\end{table}
The $c_i$ in the expressions of the polynomials are some constant parameters (that can also be vanishing). Notice that all the block classes (a), (b), and (c) are labelled by two integer parameters $r$ and $t$. In the following, we will refer to a given block in some class using the notation $\mathcal{B}_{(i)}$, with $i=a,b,c$, suppressing the dependence on $r$ and $t$ for graphical ease. For the explicit expressions of the blocks concretely realizing these polynomials, we refer to \nameref{Appendix A}. In addition, the minimal subalgebras $\mathfrak{s}_{\text{min}}$ in which the blocks $\mathcal{B}_{(i)}$ live are a natural generalization of the subalgebras $\mathfrak{u}(p)$ in which the blocks $\mathcal{A}_s$ \eqref{Higgs block Ak} in the $(A_j,A_l)$ case resided.\\

\indent It is useful at this point, in order to summarize the above argument, to fully restate the recipe to analyze the M-theory dynamics on the $(A_k,D_n)$ singularities:
\begin{enumerate}\label{list}
\item Choose a $(A_k,D_n)$ theory and compute its brane locus using equation (\ref{deformed Dn 2}).
\item Factorize the brane locus into factors corresponding to irreducible blocks, listed in table \eqref{Table 1}.
\item Build the Higgs field $\Phi$ corresponding to the theory $(A_k,D_n)$ direct-summing the irreducible blocks found at the previous point. The result is a Higgs $\Phi$ in the shape \eqref{Higgs D4} made up only of irreducible blocks.
\item Compute the stabilizer of such Higgs field, obtaining the continuous and discrete flavour/gauge symmetries.
\item Compute the matter modes localized near the branes intersections, as well as their charges under the flavour/gauge group.
\end{enumerate}

Notice that the main difference with respect to the  $(A_j,A_l)$ series lies at the second point of the recipe: it is crucial to decompose the brane locus of the $(A_k,D_n)$ singularities into irreducible blocks, as intended in table \eqref{Table 1}.\\
\indent On the other hand, in a completely analogous way with respect to the $(A_j,A_l)$ singularities, the continuous and discrete groups that are preserved by the deformed D6 brane stack can be obtained by computing the stabilizer of the Higgs field, made up of the blocks in table \ref{Table 1}.\\
\indent In the generic case, valid for \textit{all} the  $(A_k,D_n)$ singularities, the Higgs field is never made up of more than one copy of the exact same block: this means that, if more than one block of type $\mathcal{B}_{(i)}$ enters the Higgs (say e.g. $\mathcal{B}_{(i)}^{(1)}$ and $\mathcal{B}_{(i)}^{(2)}$), then they either have different sizes or they possess different constant coefficients. In this case the preserved flavour (continuous) and gauge (discrete) group is nothing but the direct product of the centers of the groups $\mathfrak{S}_{\text{min}}$ whose Lie algebras are the minimal subalgebras $\mathfrak{s}_{\text{min}}$ in which the blocks reside\footnote{In general, the Lie algebra $\mathfrak{s}_{\text{min}}$ is not enough to determine the global structure of the group $\mathfrak{S}_{\text{min}}$. In our case, however, we can use the fact that $\mathfrak{S}_{\text{min}}$ must be a subgroup of the 7d gauge group to fix the global structure.}. In other words, for a Higgs as in \eqref{Higgs D4}, given the subgroup $\mathfrak{S}_{\text{min},j}$ corresponding to the minimal subalgebra $\mathfrak{s}_{\text{min},j}$ in which the block $\mathcal{B}^{(j)}$ lives (we momentarily suppress the lower index labelling the type of block, which can be any), we have:
\begin{equation}
    \label{center of the minimal subalgebras}
    \text{Stab}(\Phi) = Z(\mathfrak{S}_{\text{min},1}) \times ... \times Z(\mathfrak{S}_{\text{min},m}), 
\end{equation}
with $m$ the number of blocks appearing in the Higgs field $\Phi$.
As a result, we can easily rewrite Table \ref{Table 1} explicitly stating the center of the subgroup $\mathfrak{S}_{\text{min}}$ corresponding to each block, yielding:
\renewcommand{\arraystretch}{1.5}
\begin{table}[H]
\begin{center}
\begin{tabular}{|c|c|c|c|}
\hline
\textbf{$\mathcal{B}$}&   $\boldsymbol{\chi(\mathcal{B})}$ & $\boldsymbol{\mathfrak{s}_{\text{min}}}$ & $\boldsymbol{Z(\mathcal{\mathfrak{S}_{\text{min}}})}$\\
\hline
(a) & $\left(\xi ^{2 r+1}+c_1 w^t\right) \left(\xi ^{2 r+1}-c_1 w^t\right)$ & $\mathfrak{su}(2r+1)$  & $U(1)$  \\
\hline
(b) &$\xi^2 \left(\xi ^{2 r}+c_2 w^{2 t+1}\right)$& $\mathfrak{so}(2r+2)$  & $\mathbb{Z}_2$ \\
\hline
(c) & \begin{tabular}{c}$\xi^{4 r}+c_3 w^{2t+1}\xi ^{2 n}+c_4 w^{2(2t+1)}$ \\
 $(r,2t+1)$ coprime
		\end{tabular} & $\mathfrak{so}(4r)$ & $\mathbb{Z}_2$ \\
\hline
\end{tabular}
\caption{Irreducible blocks and stabilizer groups.}
\label{Table 2}
\end{center}
\end{table}
Letting $2u$ be the size of the matrix representation of the minimal subalgebra $\mathfrak{s}_{\text{min}}$ in which a generic block lives, according to table \ref{Table 1}, then the explicit realizations of the generators of the centers $Z(\mathfrak{S}_{\text{min}})$ are:
\begin{equation}\label{explicit groups}
    U(1) = \left(\begin{array}{cc}
       e^{i\alpha}\mathbbm{1}_u  & 0  \\
     0    & e^{-i\alpha}\mathbbm{1}_u \\
    \end{array} \right), \hspace{1cm} \mathbb{Z}_2 = \left(\pm \mathbbm{1}_{2u} \right).
\end{equation}
Remarkably, the above table furnishes a powerful tool to analyze the resolutions of $(A_k,D_n)$ singularities: if the irreducible polynomial factorization \eqref{Higgs D4} of a given singularity contains at least one block of type (a), then it admits a small resolution inflating a 2-cycle. The number of blocks of type (a) predicts the maximum number of 2-cycles that can be resolved. This happens because the $U(1)$ groups preserved by blocks of type (a) come from the reduction of the M-theory 3-form $C_3$ on the 2-cycles inflated by the resolution.\\
\indent Furthermore, we can give a useful criterion to predict how many uncharged hypers (both under the flavour and gauge groups) are localized at the intersection of the D6 branes, just by taking a look at the irreducible block-decomposition of the brane locus. We have previously seen that the flavour and gauge groups in the 5d theory are determined by the decomposition of the brane locus into irreducible blocks $\mathcal{B}_{(i)}$ of the classification in Table \ref{Table 2}. The discrete gauge groups can be explicitly realized as the diagonal matrices \eqref{explicit groups}, that act non-trivially only on modes localized in the off-diagonal blocks. Analogously, only the modes in the off-diagonal blocks can be charged under the $U(1)$ flavour groups in \eqref{explicit groups}, except for the charge 2 modes, that are always localized inside the blocks of type (a). \\
\indent As a result the uncharged localized hypers w.r.t.\ the discrete gauging can be found only inside the blocks $\mathcal{B}_{(i)}$. In pictures, this means that we can have uncharged hypers only inside the minimal subalgebras $\mathfrak{s}_{\text{min}}$ from Table \ref{Table 2}:
\begin{equation}
    \left(\begin{array}{c|c|c}
\mathcal{B}_{(i)} & & \\
 \hline
 & \mathcal{B}_{(j)}& \\
 \hline
  & &\mathcal{B}_{(l)} \\
    \end{array} \right) \longrightarrow \left(\begin{array}{c|c|c}
uncharged &\textcolor{red}{charged} &\textcolor{red}{charged} \\
 \hline
\textcolor{red}{charged} & uncharged& \textcolor{red}{charged}\\
 \hline
 \textcolor{red}{charged} &\textcolor{red}{charged} &uncharged \\
    \end{array} \right).
\end{equation}

\indent We can explicitly summarize the number of uncharged hypers under the discrete groups appearing in each block $\mathcal{B}_{(i)}$ using their dependence on the parameters $r$ and $t$ in Table \ref{Table 2}:
\renewcommand{\arraystretch}{1.5}
\begin{table}[H]
\begin{center}
\begin{tabular}{|c|c|c|}
\hline
\multirow{2}{*}{$\boldsymbol{\mathcal{B}}$}&  \multicolumn{2}{c|}{\textbf{Uncharged hypers}} \\
\cline{2-3}
& $\boldsymbol{t=0}$  & $\boldsymbol{t\geq 1} $  \\
\hline
(a) $=\left(\xi ^{2 r+1}+c_1 w^t\right) \left(\xi ^{2 r+1}-c_1 w^t\right)$ & \multicolumn{2}{c|}{$\frac{t(t-1)}{2}$}\\
\hline
(b) $=\xi ^2 \left(\xi ^{2 r}+c_4 w^{2 t+1}\right)$  & \multicolumn{2}{c|}{$2t(r+1)$}\\
\hline
(c) $=\xi ^{4 r}+c_2 w^{2t+1} \xi ^{2 n}+c_3 w^{2(2t+1)}$ &  $r-1$ &  $4 t^2+2 r-1$ \\
\hline
\end{tabular}
\caption{Irreducible blocks and uncharged hypers under discrete symmetries.}
\end{center}
\end{table}
\vspace{0.3cm}

\textit{Let us summarize what we have shown so far: just by looking at the  brane locus factorization of a $(A_k,D_n)$ singularity we are able to predict the allowed resolution, the flavour and gauge groups and the number of uncharged hypers w.r.t\ the discrete groups in the 5d theory. In addition, by performing easy mechanical computations, we can compute all the 5d modes and their respective charges under the flavour and gauge groups, completely  characterizing the Higgs Branch.}\\
\indent We have done this explicitly for all the $(A_k,D_n)$ singularities for $k=1,\ldots 8$ and $n=4,\ldots 15$, reporting the results for the dimension of the Higgs Branches of the 5d theories, the continuous and discrete symmetries, as well as the charges of the localized modes, in \nameref{Appendix C}.\\
\indent In this regard, it is interesting to notice that there is a connection between the discrete symmetries enjoyed by the 5d theory and the one-form symmetry of the Argyres-Douglas theories arising from the geometric engineering of Type IIB theory on the $(A_k,D_n)$ singularities, as computed by \cite{Closset:2020scj}. More specifically we found that, given a starting 7d gauge group $SO(2n)$, the discrete symmetries in 5d quotiented by the center of $SO(2n)$ are exactly equal to the one-form symmetries of 4d Argyres-Douglas theories. In other words, given a  $(A_k,D_n)$ singularity, we define as $G_{\text{gauge}}^0$ the discrete gauge symmetry of the 5d theory quotiented by the centre of $SO(2n)$, i.e.\:
\begin{equation}\label{center torsion}
    G_{\text{gauge}}^0 = \frac{G_{\text{gauge}}}{Z(SO(2n))}.
\end{equation}
Then we find that the groups $G_{\text{gauge}}^0$ are:
\begin{table}[H]
\begin{center}
\renewcommand{\arraystretch}{0.9}
\begin{tabular}{|c||cccccccccccc|}
\hline
$G_{\text{gauge}}^0$ & $D_{4}$ & $D_{5}$ &$ D_{6}$ & $D_{7}$ & $D_{8}$ & $D_{9}$ & $D_{10}$ & $D_{11}$ & $D_{12}$ & $D_{13}$ &$ D_{14}$ &$ D_{15}$ \\
\hline \hline $A_{1}$ & 0 & 0 & 0 & 0 & 0 & 0 & 0 & 0 & 0 & 0 & 0 & 0 \\
$A_{2}$ & $\mathbb{Z}_{2}$ & 0 & 0 & $\mathbb{Z}_{2} $& 0 & 0 & $\mathbb{Z}_{2}$ & 0 & 0 & $\mathbb{Z}_{2}$ & 0 & 0 \\
$A_{3}$ & 0 & $\mathbb{Z}_{2}$ & 0 & 0 & 0 & $\mathbb{Z}_{2}$& 0 & 0 & 0 & $\mathbb{Z}_{2}$ & 0 & 0 \\
$A_{4} $& 0 & 0 & $\mathbb{Z}_{2}^{2}$ & 0 & 0 & 0 & 0 & $\mathbb{Z}_{2}^{2}$ & 0 & 0 & 0 & 0 \\
$A_{5}$ & 0 & 0 & 0 & $\mathbb{Z}_{2}^{2}$ & 0 & 0 & 0 & 0 & 0 & $\mathbb{Z}_{2}^{2}$ & 0 & 0 \\
$A_{6}$ & 0 & 0 & 0 & 0 & $\mathbb{Z}_{2}^{3}$ & 0 & 0 & 0 & 0 & 0 & 0 & $\mathbb{Z}_{2}^{3}$ \\
$A_{7}$ & 0 & 0 & 0 & 0 & 0 & $\mathbb{Z}_{2}^{3}$ & 0 & 0 & 0 & 0 & 0 & 0 \\
$A_{8}$ & $\mathbb{Z}_{2}$ & 0 & 0 & $\mathbb{Z}_{2}$ & 0 & 0 & $\mathbb{Z}_{2}^{4}$ & 0 & 0 & $\mathbb{Z}_{2}$ & 0 & 0 \\
\hline
\end{tabular}
\caption{Discrete gauge groups of $(A_k,D_n)$ theories modded by the center of $SO(2n)$.}
\label{Table 5}
\end{center}
\end{table}
\vspace{0.3cm}
Notice that Table \ref{Table 5} is manifestly identical to the table presented in \cite{Closset:2020scj}.\\
\indent This confirms the expectation of \cite{Closset:2020scj,Closset:2020afy} that 1-form symmetries of Type IIB reduced on $(A_k,D_n)$ singularities are linked to 0-form discrete symmetries of the 5d theories engineered with M-theory on the same singularities. We note that in equation \eqref{center torsion}, where we have quotiented by the center of the 7d group, we have apparently made an arbitrary choice of global structure of the 7d theory we started with. The M-theory interpretation of this ambiguity was deepened in \cite{Albertini:hc,Garcia-Etxebarria:rw}. We will address this point more thoroughly in a forthcoming paper.\\

\indent In the next sections we concretely apply the machinery we have set up to study M-theory on the $(A_k,D_n)$ singularities, exhibiting explicit examples admitting no resolution, an infinite family displaying one resolved 2-cycle, and a final example with two resolved 2-cycles.

\subsection{Example 1: no resolution}\label{no resolution section}
In this section we tackle the $(A_2,D_4)$ singularity, the simplest non-trivial singularity of type $(A_k,D_n)$ that admits no resolution, fully characterizing its Higgs branch. Its defining equation is:
\begin{equation}
     x^2+zy^2+z^{3}+w^{3}=0, \quad \quad (x,y,w,z) \in \mathbb C^4.
\end{equation}
\indent To complete this task, we follow the recipe outlined in the preceding section: the starting point is the brane locus, that can be computed employing equation \eqref{deformed Dn 2}. It is immediate to see that the result is:
\begin{equation}
\Delta_{(A_2,D_4)} = \xi ^2(\xi^6+ w^3).
\end{equation}
We can now completely factorize it into the irreducible blocks in table \ref{Table 1}, obtaining:
\begin{equation}
\Delta_{(A_2,D_4)}= \underbrace{\xi ^2 \left(\xi ^2+w\right)}_{\text{type (b)}} \underbrace{\left(\xi ^4+w^2-\xi ^2 w\right)}_{\text{type (c)}},
\end{equation}
where we have highlighted the specific type of blocks. The fact that there are no blocks of type (a), corresponding to $U(1)$ flavour groups, indicates that the singularity is non-resolvable. Direct summing the irreducible blocks we obtain the explicit Higgs field reproducing the D6 brane configuration of the $(A_2,D_4)$ theory, where each diagonal block is in the basis \eqref{so basis}:
\begin{equation}\label{non resolvable higgs}
\Phi = \left(
\begin{array}{cccc|cccc}
 0 & 1 & 0 & \frac{w}{4} & 0 & 0 & 0 & 0 \\
 -\frac{w}{4} & 0 & -\frac{w}{4} & 0 & 0 & 0 & 0 & 0 \\
 0 & 1 & 0 & \frac{w}{4} & 0 & 0 & 0 & 0 \\
 -1 & 0 & -1 & 0 & 0 & 0 & 0 & 0 \\
 \hline
 0 & 0 & 0 & 0 & 0 & 1 & 0 & \frac{w}{4} \\
 0 & 0 & 0 & 0 & \frac{3 w}{4} & 0 & -\frac{w}{4} & 0 \\
 0 & 0 & 0 & 0 & 0 & 1 & 0 & -\frac{3w}{4}  \\
 0 & 0 & 0 & 0 & -1 & 0 & -1 & 0 \\
\end{array}
\right).
\end{equation}
We decompose $\Phi$ as:
\begin{equation*}
   \Phi = \left(\begin{array}{c|c}
  \mathcal{B}_{(b)}   & 0 \\
  \hline
  0   &  \mathcal{B}_{(c)} 
\end{array} \right) \quad\quad \text{with: } 
\begin{cases}
\chi(\mathcal{B}_{(b)}) = \xi ^2 \left(\xi^2+w\right) \quad\quad\quad\hspace{0.1cm} \text{type (b)}, \\
\chi(\mathcal{B}_{(c)}) =\left(\xi ^4+w^2-\xi ^2 w\right) \quad \text{type (c)},\\
\end{cases}
\end{equation*}
where we have explicitly highlighted the block decomposition. It is also easy to verify that the relationship \eqref{deformed Dn 2} between the Higgs field and the algebraic definition of the $(A_2,D_4)$ precisely holds.\\
\indent The stabilizer group of \eqref{non resolvable higgs}, corresponding to the flavour (for the continuous part) and gauge (for the discrete part) symmetry of the 5d SCFT can be promptly read off Table \ref{Table 2}, noticing that each block contributes a $\mathbb{Z}_2$ factor, yielding:
\begin{equation}
    G_{\text{gauge}} = \mathbb{Z}_2
\end{equation}
and a trivial flavour group $G_{\text{flavour}}$. 
We have only one factor of  $\mathbb{Z_2}$ acting on the modes, as opposed to the full stabilizer of $\Phi$, which is $\mathbb{Z_2}\times\mathbb{Z_2}$, as the diagonal combination belongs to the center of the seven-dimensional gauge group $SO(8)$.\\
\indent Studying fluctuations $\varphi$ around the Higgs background $\Phi$, subject to the equivalence $\varphi \sim \varphi+\left[\Phi,g\right]$, with $g$ a generic matrix of parameters in $\mathfrak{so}(8)$, we can identify the content of the Higgs branch, obtaining:
\begin{equation}
  \varphi = \left(\begin{array}{c|c}
 \emptyset   & 4 \text{ modes} \\
  \hline
  4 \text{ modes}   &  \emptyset
\end{array} \right) .
\end{equation}
All in all, we get a total of 4 hypers, as expected from previous results in the literature \cite{Xie} \cite{Carta Mininno}.\\
\indent Summarizing, we find that the Higgs branch of the $(A_2,D_4)$ theory coincides with existing results \cite{Closset:2020scj}:
\begin{equation}
    \text{HB}_{(A_2,D_4)} = \frac{\mathbb{C}^8}{\mathbb{Z}_2},
\end{equation}
with $\mathbb Z_2$ acting reflecting all the coordinates of $\mathbb C^8$.

\subsection{Example 2: one resolved 2-cycle}\label{example Closset}
In this section we get to a more interesting family of examples, admitting a small resolution of a single 2-cycle, a fact that is signalled by the appearance of a $U(1)$ symmetry in the stabilizer of the Higgs background. The family is formed by the singularities $(A_{2k-1},D_{2kn+1})$:
\begin{equation}\label{Closset family}
    x^2+ z y^2  + z^{2kn}+ w^{2k} =0, \quad \quad (x,y,w,z) \in \mathbb C^4.
\end{equation}
Such family was pinpointed and studied by Closset et al.\footnote{C. Closset, private communication with Andrés Collinucci.}, and employing our techniques we show how to fully characterize their Higgs branch.\\
\indent The D6 brane loci corresponding to the geometry \eqref{Closset family} can be computed as:
\begin{equation}\label{brane locus Closset}
 \Delta(\xi^2,w) = \xi^2(\xi^{4kn}+w^{2k}) .  
\end{equation}
Fully decomposing the brane locus into factors of the allowed form, presented in Table \ref{Table 2}, we get:
\begin{equation}\label{reduced closset}
   \Delta(\xi^2,w) = \underbrace{\xi^2}_{\text{type (a)}}\prod_{s=0}^{k-1}\underbrace{(\xi^{4n}+e^{2\pi is/k}w^2)}_{\text{type (c)}}.
\end{equation}
The fact that we obtain a block of type (a), that preserves a $U(1)$ flavour symmetry, is the telltale sign that a small resolution of a single 2-cycle is allowed by this Higgs configuration.\\
\indent Direct-summing the blocks corresponding to each of the factors in \eqref{reduced closset} we obtain the full Higgs field $\Phi$, expressed in an appropriate basis of the Lie algebra $\mathfrak{so}(4kn+2)$, where each block factor is in the basis \eqref{so basis}:
\begin{equation}\label{closset higgs}
    \Phi = \left(\begin{array}{c|ccc}
     \mathcal{B}_{(a)}    &  &  & \\
     \hline
      & \mathcal{B}_{(c)}^{(1)} & & \\
     & & \ddots & \\
     & & & \mathcal{B}_{(c)}^{(k)} \\
    \end{array} \right), \quad\quad  \text{with: } \begin{cases}
    \chi(\mathcal{B}_{(a)}) = \xi^2, \\
    \chi(\mathcal{B}_{(c)}^{(1)}) =\xi^{4n}+w^2, \\
    \quad\quad\quad \vdots \\
    \chi(\mathcal{B}_{(c)}^{(k)}) =\xi^{4n}+e^{2\pi i(k-1)/k}w^2. \\
    \end{cases}
\end{equation}
In this way, we can trivially check that the determinant of the full Higgs correctly reproduces the brane locus \eqref{brane locus Closset} of the $(A_{2k-1},D_{2kn+1})$ singularities.\\
\indent Using the techniques outlined in the previous section, namely studying fluctuations of $\Phi$, we can easily compute the hypermultiplet content of the 5d theory and their charges under the flavour and gauge symmetries.\\
\indent As regards the matter modes, they can be nicely displayed in a block form, following the structure of the Higgs field $\Phi$ \eqref{closset higgs}:
\begin{equation}\label{modes closset}
    \Phi = \left(\begin{array}{c|ccc}
     \mathcal{B}_{(a)}    &  &  & \\
     \hline
      & \mathcal{B}_{(c)}^{(1)} & & \\
     & & \ddots & \\
     & & & \mathcal{B}_{(c)}^{(k)} \\
    \end{array} \right) \longrightarrow 
   \text{modes} = \left(\begin{array}{c||c|c|c|c}
      \emptyset & \textcolor{red}{2} & \textcolor{red}{\cdots} & \textcolor{red}{\cdots} & \textcolor{red}{2}\\
      \hline
      \hline
      \textcolor{red}{2} & 2n-2  & \textcolor{ForestGreen}{4n} & \textcolor{ForestGreen}{\cdots} & \textcolor{ForestGreen}{4n} \\
      \hline
      \textcolor{red}{\vdots}& \textcolor{ForestGreen}{4n} & \ddots & \textcolor{ForestGreen}{\ddots} & \textcolor{ForestGreen}{\vdots} \\
      \hline
      \textcolor{red}{\vdots}& \textcolor{ForestGreen}{\vdots} &\textcolor{ForestGreen}{\ddots} & \ddots & \textcolor{ForestGreen}{4n} \\
      \hline
      \textcolor{red}{2} & \textcolor{ForestGreen}{4n} &\textcolor{ForestGreen}{\cdots} &\textcolor{ForestGreen}{4n} & 2n-2  \\
    \end{array} \right).
\end{equation}
The stabilizer group $G_{\text{stab}}$ of $\Phi$ reads, according to table \eqref{Table 2}:
\begin{equation}
    G_{\text{stab}} = U(1)\times\underbrace{\mathbb Z_2 \times ...\times \mathbb Z_2}_{k} = \left(\begin{array}{c|ccc}
    U(1)    &  &  & \\
     \hline
      & \mathbb{Z_2}^{(1)}  & & \\
     & & \ddots & \\
     & & & \mathbb{Z_2}^{(k)}  \\
    \end{array} \right),
\end{equation}
obtaining a $U(1)$ flavour group coming from the resolved 2-cycle, as well as $k$ discrete $\mathbb{Z}_2$ gauge groups. Taking a look at the block structure of \eqref{modes closset}, it is immediate to see that different colors correspond to different charges under the flavour $U(1)$ and the gauge $\mathbb Z_2^{(i)}$ groups:
\begin{equation}
    \begin{cases}
    \textcolor{red}{red:} \quad\quad  \hspace{0.4cm}\text{charge $\pm 1$ under $U(1)$, and one of the $\mathbb Z_{2}^{(i)}$}\\
    \textcolor{ForestGreen}{green:} \quad\quad \text{charged under two $\mathbb{Z}_2^{(i)}$} \text{ factors}\\
    \textcolor{black}{black:} \quad\quad \hspace{0.1cm}\text{uncharged} \\
    \end{cases}
\end{equation}
We remark that we will use this color notation also in the systematic tables of \nameref{Appendix C}.\\
\indent As regards the discrete charges, it is of course possible to precisely track the charge of each mode under every $\mathbb{Z}_2^{(i)}$ group, just by taking a look at the block structure \eqref{modes closset}.\\
\indent Summing up, for the infinite family of singularities $(A_{2k-1},D_{2kn+1})$ we find a total of $2nk^2+k-nk$ hypers, with charges:
\begin{itemize}
\item $2k$ hypers charged under $U(1)$ \\
\item $2nk(k-1)$ hypers charged only under some $\mathbb{Z}_2$ \\
\item $k(n-1)$ uncharged hypers \\
\end{itemize}
As result, the general formula for the Higgs Branch is:
\begin{equation}
    \text{HB} = \mathbb{C}^{2k(n-1)}\times\mathbb{C}^{4k}\times\frac{\mathbb{C}^{4nk(k-1)}}{\mathbb{Z}_2^{k-1}}.
\end{equation}
Notice that, as it happened in the previous example \eqref{no resolution section}, one combination of the $\mathbb Z_{2}^{(i)}$ is always decoupled, leaving the effective flavour/gauge group $G$ as:
\begin{equation}
    G_{\text{flavour}} = U(1),\hspace{1cm} G_{\text{gauge}} = \underbrace{\mathbb Z_2 \times ...\times \mathbb Z_2}_{k-1}.
\end{equation}
We stress that we have a complete understanding of the structure of the Higgs Branch. In other words, we can completely reconstruct, from \eqref{modes closset}, the action of the discrete group $G_{\text{gauge}}$ giving, e.g., the Hilbert series of the Higgs Branch. For example, choosing $k = 3$, $n = 1$, we get, using \cite{Benvenuti:2006qr} the Molien formula:
\begin{equation}
    \text{HS} = \frac{N(t)}{D(t)}, 
\end{equation}
with
\begin{eqnarray}
N(t) & = & \scalemath{0.8}{t^{20}-10 t^{19}+190 t^{18}-570
   t^{17}+4845 t^{16}-7752 t^{15}+38760
   t^{14}-38760 t^{13}+125970 t^{12}+} \nonumber \\
   &&
  \scalemath{0.8}{ -83980 t^{11} 
   +184756 t^{10}-83980 t^9+125970
   t^8-38760 t^7 +}\nonumber \\
   &&\scalemath{0.8}{+38760 t^6-7752 t^5+4845
   t^4-570 t^3+190 t^2-10 t+1}, \nonumber \\
   D(t)& = & \scalemath{0.8}{(t-1)^{36}
   (t+1)^{20}}. \nonumber \\
\end{eqnarray}
\subsection{Example 3: two resolved 2-cycles}
As a final example of the practicality of our techniques, we examine the $(A_{3},D_4)$ singularity, that can be readily shown to admit a small resolution inflating two 2-cycles. Its algebraic definition is:
\begin{equation}
    x^2+zy^2+z^{3}+w^{4} = 0.
\end{equation}
The corresponding brane locus reads:
\begin{equation}
    \Delta(\xi^2,w) = \xi^2(\xi^{6}+w^{4}).
\end{equation}
The factorized brane locus hence is:
\begin{equation}\label{brane locus two resolutions}
    \Delta(\xi^2,w)= \underbrace{\xi^2}_{\text{type (a)}}\underbrace{(\xi^{3}+i w^{2})(\xi^{3}-i w^{2})}_{\text{type (a)}}.
\end{equation}
Notice the appearance of two different blocks of type (a), signalling the presence of two commuting preserved $U(1)$ groups, and hence two resolvable 2-cycles.\\
\indent The irreducible brane locus \eqref{brane locus two resolutions} can be explicitly realized via the following Higgs background, in an appropriate basis of $\mathfrak{so}(8)$\footnote{The first $2\times 2$ vanishing diagonal block, as well as the remaining 6$\times$6 diagonal block, are in the standard basis \eqref{so basis}.}:
\begin{equation}
    \Phi =\scalemath{0.8}{\left(
\begin{array}{cc|ccc|ccc}
 0 & 0 & 0 & 0 & 0 & 0 & 0 & 0 \\
 0 & 0 & 0 & 0 & 0 & 0 & 0 & 0 \\
 \hline
 0 & 0 & 0 & 1 & 0 & 0 & 0 & 0 \\
 0 & 0 & 0 & 0 & \sqrt{i} w & 0 & 0 & 0 \\
 0 & 0 & \sqrt{i} w & 0 & 0 & 0 & 0 & 0 \\
 \hline
 0 & 0 & 0 & 0 & 0 & 0 & 0 & -\sqrt{i} w \\
 0 & 0 & 0 & 0 & 0 & -1 & 0 & 0 \\
 0 & 0 & 0 & 0 & 0 & 0 & -\sqrt{i} w & 0 \\
\end{array}
\right)}.
\end{equation}
Considering fluctuations $\varphi$ of the background up to gauge transformations we find the following matter modes:
\begin{equation}
    \text{modes} = \left(\begin{array}{c|c|c}
     0  & \textcolor{red}{2} & \textcolor{red}{2}\\
      \hline
   \textcolor{red}{2}      & 1& \textcolor{blue}{2}\\
   \hline
     \textcolor{red}{2}    & \textcolor{blue}{2} & 1 \\
 
    \end{array} \right).
\end{equation}
The colors indicate different charges under the flavour group\footnote{As predicted in \cite{Collinucci}, for $D_n$ families we can find five-dimensional modes of charges up to two.} $G_{\text{flavour}} = U(1)_{\alpha}\times U(1)_{\beta}$, that can be explicitly represented as:
\begin{equation*}
    G_{\text{flavour}} = \left(
\begin{array}{c|c|c}
\scalemath{0.7}{\begin{array}{cc}
  e^{i\alpha}   &0  \\
    0 & e^{-i\alpha}
\end{array}}& 0 & 0 \\
\hline
0 & e^{i\beta}\mathbbm{1}_3 & 0 \\
\hline
0 & 0 & e^{-i\beta}\mathbbm{1}_3\\
\end{array}
\right) \Longrightarrow \begin{cases}
\textcolor{red}{red:} \quad\quad  \hspace{0.4cm}\text{charge $\pm 1$ under $U(1)_{\alpha}$ and $U(1)_{\beta}$}\\
    \textcolor{blue}{blue:} \quad\quad \hspace{0.3cm}\text{charge $\pm 2$ under $U(1)_{\beta}$} \\
    \textcolor{black}{black:} \quad\quad \hspace{0.1cm}\text{uncharged} \\
\end{cases} 
\end{equation*}
As a result, we find a total of 7 hypers, charged as:
\begin{itemize}
    \item 4 hypers with charge 1
    \item 2 hypers with charge 2
    \item 1 uncharged hyper
\end{itemize}
The Higgs Branch hence is:
\begin{equation}
    \text{HB} = \mathbb{C}^2 \times\mathbb{C}^{12},
\end{equation}
where the $\mathbb{C}^2$ factor refers to the uncharged hyper, and the $\mathbb{C}^{12}$ factor to the hypers charged under the flavour group.

\section{T-branes hierarchy}
\label{T-branes section}
As we have briefly mentioned in the case of $(A_j,A_l)$ singularities, the choice of the Higgs background $\Phi$ deforming the stack of D6 branes is not unique. As first realized by \cite{Cecotti}, and further investigated in a deluge of subsequent works \cite{Marchesano:2020idg,Hassler:2019eso,Barbosa:2019bgh,Bena:2019rth,Marchesano:2019azf,Carta:2018qke,Marchesano:2017kke,Collinucci:2017bwv,Bena:2017jhm,Anderson:2017rpr,Mekareeya:2016yal,Marchesano:2016cqg,Bena:2016oqr,Collinucci:2016hpz,Cicoli:2015ylx,Collinucci:2014taa,Collinucci:2014qfa,Anderson:2013rka,Chiou:2011js} there exist various possible Higgs backgrounds dual to M-theory on the same geometry, but displaying a different physical content. In our cases, this means that in general many different Higgs fields $\Phi^{(i)}(w)$ realizing the brane-loci of the $(A_j,A_l)$ singularities are allowed. These various choices can be neatly labelled using the Lie algebra formalism involving nilpotent orbits: given a Higgs field $\Phi^{(i)}(w)$, we define as $\Phi^{(i)}_0 = \Phi^{(i)}(w=0)$ its constant component. We found that $\Phi_{0}^{(i)}$ is always nilpotent for all the $(A_j,A_l)$ and $(A_k,D_n)$, and thus belongs to some nilpotent orbit $\mathcal{O}_0^{(i)}$ of $A_j$ (supposing $j>k$). Consequently we can label every Higgs $\Phi^{(i)}(w)$ using the nilpotent orbit $\mathcal{O}_0^{(i)}$ in which its constant component resides. Furthermore, the codimension formula \eqref{codimension formula} relates the nilpotent orbit $\mathcal{O}_0^{(i)}$ to the number of linearly independent elements of the 7d gauge algebra that support 5d modes localized at the intersection of the D6 branes. For the $(A_j,A_l)$ singularities the story ends here: in order to obtain the Higgs background for $(A_j,A_l)$ yielding the maximal number of modes, we take the blocks in $\mathcal{A}^{(i)}$ in \eqref{Higgs block Ak}, evaluated on $w=0$, to lie in the biggest-codimension nilpotent orbit $\mathcal{O}^{(i)}_0$ compatible with the geometry, namely reproducing the brane locus.\\
\indent The exact same phenomenon happens in the $(A_k,D_n)$ singularities, although the hierarchy of the different Higgs backgrounds is more  complicated. The goal of this section is to show how a classification of the allowed Higgs backgrounds is possible, providing an explicit example.\\

\indent The starting point, as always, is the brane locus. The only constraint that must be imposed on the Higgs $\Phi(w)$ is (\ref{deformed Dn 2}), that we reproduce here for convenience:
\begin{equation}\label{brane locus}
\text{det }(\xi \mathbb{1}+\Phi(w)) = \Delta(\xi^2,w) = \xi^2(\xi^{2n-2}+w^{k}).
\end{equation}

As we have said, there is vast space for ambiguities in the choice of the Higgs, giving rise to a hierarchy governed by the nilpotent orbits that can be associated to the Higgs itself. Let us see how this precisely comes about.\\
\indent Generally speaking, each Higgs comprises constant entries, along with entries depending on $w$ ($w$-entries).\\
\indent Correspondingly, by considering the constant and $w$-entries separately, we can analyze their orbit structure. In particular, for all the cases in (\ref{brane locus}), we now show how to associate both the constant entries and the $w$-entries to \textbf{nilpotent orbits}, that can be classified by suitable partitions of $[2n]$ as the Higgs $\Phi$ lives in the algebra $\mathfrak{so}(2n)$. As is well known in the mathematical literature, nilpotent orbits are organized hierarchically along Hasse diagrams, and this structure will be reflected in the possible choices for the Higgs background, giving rise in general to different spectrums. Following the notation for the $(A_j,A_l)$ cases, we will denote the nilpotent orbit associated to the constant entries as $\mathcal{O}_0$, and the one related to the $w$-entries as $\mathcal{O}_w$. More precisely, we define:
\begin{equation}
\begin{split}
    &\mathcal{O}_0 = \text{nilpotent orbit in which }\Phi(0)\text{ lives,}\\
    &\mathcal{O}_w = \text{nilpotent orbit in which }\Phi-\Phi(0)\text{ lives.}\\
    \end{split}
\end{equation}\\
Consequently, the full Higgs field $\Phi$ can be decomposed as:
\begin{equation}
    \Phi = \Phi(0) +(\Phi-\Phi(0)) \equiv \Phi_0+\Phi_w,
\end{equation}
where $\Phi_0 \in \mathcal{O}_0$ and $\Phi_w \in \mathcal{O}_w$.\\
\indent When trying to pick a choice for $\Phi$ satisfying (\ref{brane locus}) for a given brane locus related to some $(A_{k},D_{n})$ singularity, one is confronted with the following logical steps:
\begin{itemize}
\item In general, each brane locus is compatible with many choices of $\mathcal{O}_0$\footnote{Here by compatible we mean that we can build an Higgs field $\Phi$ with $\Phi_0$ belonging to $\mathcal O_0$.}, thus giving rise to an ambiguity. There is always a \textit{minimal $\mathcal{O}_0$}, giving rise to the largest spectrum. Mathematically this is the lowest-lying orbit, among the compatible ones, in the Hasse diagram.\\
Most notably, \textit{the choice of $\mathcal{O}_0$ completely fixes the number of linearly independent elements inside the 7d gauge algebra supporting 5d localized modes}, according to the codimension formula \eqref{codimension formula}.
\item In general, each $\mathcal{O}_0$ is compatible with many \textit{bottom} orbits $\mathcal{O}_w$, namely with many different choices of $w$-entries, barely sufficient to reproduce the correct brane locus (where ``barely'' means that no ``$w$'' entry can be removed without affecting the brane locus). Among the bottom orbits $\mathcal{O}_w$ there is always a minimal $\mathcal{O}_w$, lying at the lowest position in the Hasse diagram, giving rise to the maximal number of modes.\\
Each bottom $\mathcal{O}_w$ gives rise, in general, to a different number of total 5d modes.
\item By deforming each bottom $\mathcal{O}_w$, tuning zero-entries into $w$-entries while keeping the brane locus and $\mathcal{O}_0$ fixed, we find a \textit{tower} of allowed $\mathcal{O}_w$, starting from the bottom one and terminating on a top one (there always is a top orbit, as the size of the Higgs is fixed by the brane locus).\\
Most importantly, \textit{each $\mathcal{O}_w$ belonging to the same tower\footnote{We stress that this means that the Higgs associated to the $\mathcal O_w$ in the tower is obtained turning on some $w$-entries in the Higgs associated to the bottom orbit, without modifying its brane locus.} gives rise to the same number of total modes}. In addition, towers starting from different bottom $\mathcal{O}_w$ need not be disjoint (meaning that the same $\mathcal{O}_w$ can appear in many different towers, producing different amounts of modes. What counts for the number of modes is the \textit{bottom} $\mathcal{O}_w$ at the base of the tower).
\end{itemize}

Summing up, given a brane locus in the $(A_k,D_n)$ series, a choice of the Higgs is completely determined once one picks:
\begin{equation}\label{nilpotent data}
  \begin{cases}
\text{a nilpotent orbit }\mathcal{O}_0, \text{ corresponding to the constant entries of $\Phi$}, \\
\text{a bottom orbit }\mathcal{O}_w, \text{ corresponding to the $w$-entries of $\Phi$}.\\
\end{cases}
\end{equation}

\indent In order to understand this hierarchy of choices in a more intuitive way, it is instructive to depict it graphically, indicating with segments the possible choices, and with arrows the nilpotent orbits hierarchy in the Hasse diagram sense. Notice that we have explicitly indicated the minimal $\mathcal{O}_0$ and minimal $\mathcal{O}_w$ orbits, that when combined in the choice of the Higgs yield the M-theory dynamics with the maximal number of modes. In an extensive case by case analysis we have always found that such choice is unique, but we cannot rule out the possibility that there is more than one minimal choice of $\mathcal{O}_0$  and $\mathcal{O}_w$  yielding the maximal number of modes, as there could be more than one orbit on the same level of the Hasse diagram hierarchy. We finally stress that each bottom orbit $\mathcal{O}_w$ in the picture is the starting point of a tower of orbits, obtained deforming the Higgs configuration corresponding to the bottom orbit, with the same number of total 5d modes as the ones given by the bottom orbit. We have omitted such towers for a better graphical depiction.\\
\indent Finally, notice that for every choice of $\mathcal{O}_0$ we have indicated the total number of 7d elements supporting localized 5d hypers (namely, the number of 7d elements supporting localized 5d modes given by the codimension formula \eqref{codimension formula} is \textit{twice} the number we have indicated), and that for every bottom $\mathcal{O}_w$ we have highlighted the total number of hypers.

\newpage
\begin{center}
\includegraphics[scale=0.33]{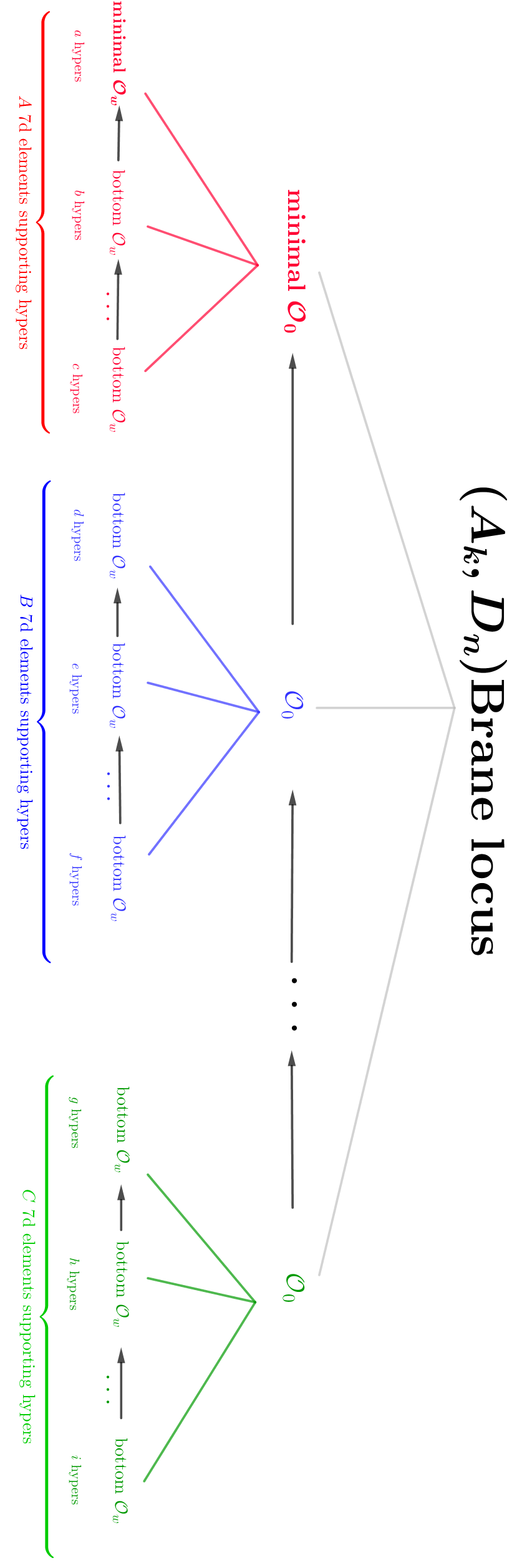}
\end{center}

\newpage
Let us now examine a concrete example, so as to make the abstract remarks above a bit more grounded. An interesting instance of brane locus giving rise to a T-brane hierarchy is $(A_8,D_8)$, that displays a remarkable structure. This singularity is non-resolvable and its brane locus is:
\begin{equation}
\Delta(\xi^2,w)=\underbrace{\xi^2(\xi^{14}+w^{9})}_{\text{type (b)}} = 0.
\end{equation}
\indent In the following picture, the red color refers to the \textcolor{red}{$\mathcal{O}_0$}, the blue color to \textcolor{blue}{bottom $\mathcal{O}_w$} and the dark arrows to dominance in the Hasse diagram sense. We have instead omitted towers with the same number of total hypers for the sake of graphical clarity. As before, we have indicated the total number of 7d elements supporting localized 5d hypers for every choice of $\mathcal{O}_0$ in the hierarchy, as well as the total number of hypers for every bottom $\mathcal{O}_w$. As it can be seen from the picture, the M-theory dynamics with maximal modes is reproduced by the lowest $\mathcal{O}_0$ with the lowest $\mathcal{O}_w$ in the Hasse diagram, yielding 32 total hypers. All the other partitions are instead T-brane configurations with a lower amount of modes.

\begin{center}
\includegraphics[scale=0.115]{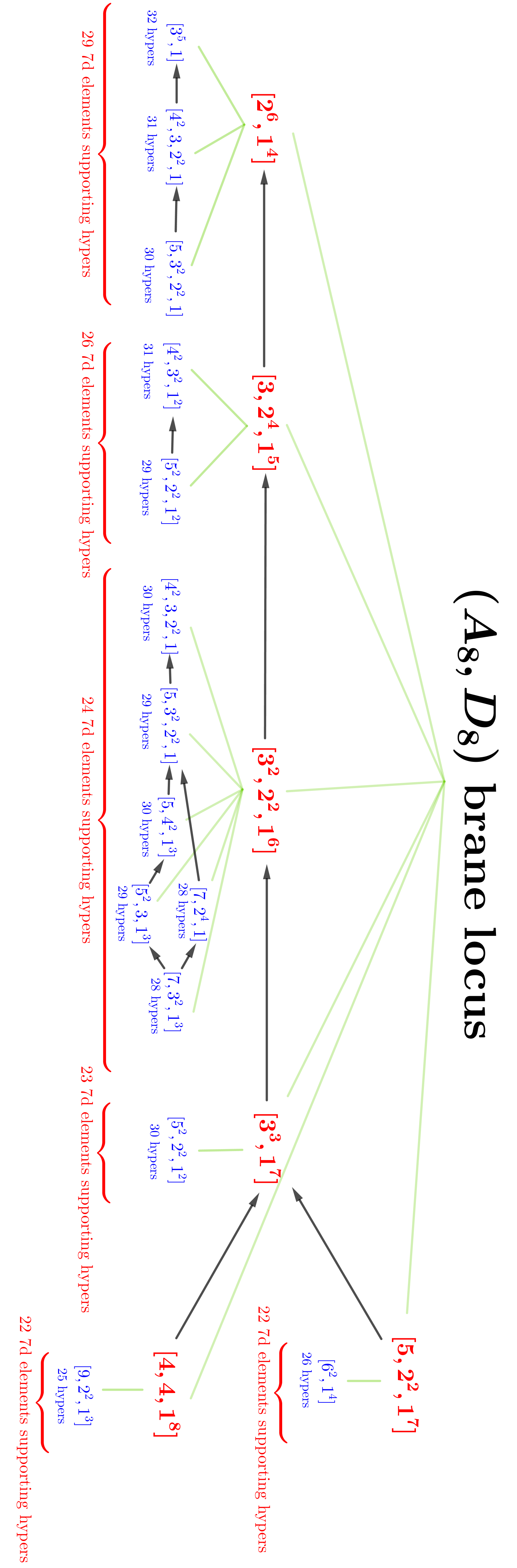}
\end{center}
It is clear that the T-brane hierarchy can be extremely rich, expressing peak complexity in the $(A_k,D_n)$ singularities and giving rise to a plethora of different Higgs backgrounds encoding the same geometry, but a different amount of localized modes. The T-brane hierarchy is of course a type IIA theory feature, and besides its interconnected structure, it is suggestive of a corresponding intricacy in the dual M-theory description. In this regard, the role and the characteristics of T-branes in M-theory have been discussed in \cite{Anderson:2013rka,Collinucci:2014taa,Collinucci:2016hpz,Collinucci:2017bwv}, showing that geometry is not enough to characterize a T-brane background, and that additional structure is needed.\\
\indent This is exactly what we have shown from a type IIA perspective in the preceding pages: the choice of the Higgs background is intrinsically ambiguous and additional non-geometric data, e.g.\ the orbits $\mathcal{O}_0$ and $\mathcal{O}_w$ in \eqref{nilpotent data}, must be specified for a full characterization of the spectrum and of the preserved symmetries.\\
\indent

\section{Conclusions}
\label{section: conclusions}
In this paper we studied the Higgs Branches of five-dimensional rank-zero SCFTs engineered by M-theory on threefold singularities of type $(A_j,A_l)$ and $(A_k,D_n)$. Our method applies to all the studied cases. We gave a closed formula \eqref{HB general expression AjAk} for all the Higgs Branches of the $(A_j,A_l)$ series, and tables for the $(A_k,D_n)$ series (prescribing, in this case, how to perform the computation in the general case). We confirm the lore that these theories engineer either discrete gaugings of hypermultiplets or free hypermultiplets, both in singularities admitting a small crepant resolution and in the so-called ``terminal singularities'' (isolated threefold singularities that do not admit any crepant K\"ahler resolution).\\

\indent  Furthermore, we clarified the open-string viewpoint on the T-branes states: open-string exotic bound states whose brane locus coincides with the one dictated by the M-theory/type IIA duality, but with some obstructed deformations. We re-organized these data nicely according to the Hasse diagrams of nilpotent orbits of (respectively, for the $(A_j,A_l)$ and the $(A_k,D_n)$ series) $\mathfrak g = A_l$ and $\mathfrak{g} = D_n$. Loosely speaking, the number of five-dimensional modes decreases when the nilpotent orbit that we associate to the T-brane state climbs up the Hasse diagram.\\

\indent  Our work opens up many possible further developments. First of all, it is known \cite{KatzMorrison} that any small resolution of a threefold isolated singularity can be locally modelled as an ADE fibration. Our paper rules out the possibility of rank-zero theories that are not free hypermultiplets or discrete gauging of free hypers, at least for the $A_j$ or $D_n$ fibered threefolds. We will address the $E_6, E_7, E_8$ cases in a forthcoming paper.\\
\indent Once we ruled out the rank-zero case, then, a natural continuation would be to apply our method also to higher-rank theories, whose resolution inflates an exceptional locus of complex-dimension two.   \\ 

\indent Finally, the structured hierarchy of T-branes that is manifest in our Type IIA analysis naturally suggests a corresponding intricacy in the dual M-theory description, in resonance with the existing literature \cite{Anderson:2013rka,Collinucci:2014taa,Collinucci:2016hpz,Collinucci:2017bwv} and further motivating the necessity to go beyond geometry to fully characterize a sensible M-theory brane background.

\section*{Acknowledgments} 

We would like to thank Roberto Valandro for his important contributions all along the project and for useful suggestions on the draft. We would also like to thank Andrés Collinucci for useful discussions and insightful suggestions. M. D. M. acknowledges Michele Graffeo for useful suggestions. A.S. wishes to thank Sergej Monavari for useful discussions.
A.S. acknowledges support by INFN Iniziativa Specifica ST\&FI. M.D.M. acknowledges support by INFN Iniziativa Specifica GAST.

\newpage
\section{Appendix A}\label{Appendix A}
\subsection{Explicit expressions of blocks in $\boldsymbol{(A_k,D_n)}$ singularities}
In this appendix, we give a schematic account of the blocks living in a subalgebra of $\mathfrak{so}(2n)$ yielding the polynomials appearing in Table \ref{Table 1} in the brane locus ($*$ entries representing either a constant or a term linear in $w$). Unless explicitly stated, we employ the basis (\ref{so basis}) for $\mathfrak{so}(\text{size of block})$.
\begin{itemize}
\item $\boldsymbol{P_{(a)}(\xi^2,w) = \left(\xi ^{2 r+1}+c_1 w^t\right) \left(\xi ^{2 r+1}-c_1 w^t\right)}$\\

The blocks $\mathcal{B}_{(a)}$ such that its characteristic polynomial (indicated with ``$\chi$'') satisfies $\chi(\mathcal{B}_{(a)}) = P_{(a)}(\xi^2,w)$ are of the form:
\begin{equation}\label{su subalgebra block}
\mathcal{B}_{(a)} = \left(\begin{array}{cc} 
\mathcal{A}_{(2r+1)\times (2r+1)} &  0 \\
0 & -\mathcal{A}_{(2r+1)\times (2r+1)}^t \\
\end{array}\right),
\end{equation}
where $\mathcal{A}_{(2r+1)\times (2r+1)}$ has the form:
\begin{equation}\label{A blocks}
\mathcal{A}_{(2r+1)\times (2r+1)} = \left(\begin{array}{CCCCCC}
0 & * & 0 & \cdots & \cdots & 0 \\
\vdots & 0 & * & 0 &  & \vdots \\
\vdots &  &\ddots & \ddots & & \vdots \\
\vdots & & &\ddots & \ddots & \vdots \\
\vdots & &  &  & 0 & * \\
* & 0 & \cdots &\cdots &\cdots & 0 \\
\end{array} \right).
\end{equation}
Notice that the form \eqref{A blocks} is analogous to the one of the blocks \eqref{Higgs block Ak} in the $(A_j,A_l)$ cases, as indeed \eqref{su subalgebra block} belongs to a $\mathfrak{su}(2r+1)$ subalgebra of $\mathfrak{so}(4r+2)$.

\newpage
\item $\boldsymbol{P_{(b)}(\xi^2,w)=\xi ^2 \left(\xi ^{2 r}+c_2 w^{2 t+1}\right)}$\\

The blocks $\mathcal{B}_{(b)}$ are of the form:
\begin{equation}
\scalemath{0.9}{
\mathcal{B}_{(b)}= \left(
\begin{array}{CCCCCC|CCCCCC}
0 & & & & & & & & & & & *\\
 & \ddots & & & & & & & & & * & \\
 & & 0 & \textcolor{red}{*} & & & & & & \iddots & & \\
 & & & 0 & & & & & \iddots & & & \\
 & & & & \ddots & & & * & & & & \\
* & & & & & 0 & * & & & & & \\
 \hline
 & & & & & 0 & 0 & & & & & *\\
 & & & & 0 & * &  & \ddots & & & & \\
 & & & \iddots & * & & & & 0 & & & \\
 & & \iddots & \iddots & & & & & \textcolor{red}{*} & 0 & & \\
 & 0 & * & & & & & & & & \ddots & \\
0 & * & & & & & & & & & & 0\\ 
\end{array}
\right)},
\end{equation}
where the $\textcolor{red}{*}$ on the over-diagonal of the upper diagonal block appears on the $\frac{(r+1)}{2}^{\text{th}}$ row if $r$ is odd and on the $\frac{r}{2}^{\text{th}}$ row if $r$ is even. The other $\textcolor{red}{*}$ on the lower diagonal block appears in the corresponding entry according to the basis \eqref{so basis}.

\newpage
\item $\boldsymbol{P_{(c)}(\xi^2,w)= \xi^{4 r}+c_3 w^{2t+1}\xi ^{2 r}+c_4 w^{2(2t+1)}}$, $(r,2t+1)$ coprime \\

The blocks $\mathcal{B}_{(c)}$ are of the form:
\begin{equation}\label{blocks C}
\scalemath{0.9}{
\Huge{\mathcal{B}_{(c)}}\normalsize= \left(
\begin{array}{CCCCCC|CCCCCC}
0 & *& & & & & & & & & & *\\
 & 0& * & & & & & & & & 0 & \\
 & & \ddots & \ddots &  & & & & & \iddots & & \\
 & & & \ddots & * & & & & \iddots & & & \\
 & & & & 0 & * & & 0 & & & & \\
* & & & & & 0 & * & & & & & \\
 \hline
0 & * & & & & & 0 & & & & & *\\
* & & & & & & * & 0& & & & \\
 & & & & & & &*&\ddots & & & \\
 & & & & & & & & \ddots & \ddots & & \\
 & & & & & & & & &* & 0 & \\
 & & & & & & & & & &* & 0\\ 
\end{array}
\right)}.
\end{equation}

\end{itemize}

\newpage

\section{Appendix B}
\label{appendix B}
\subsection{Codimension formula, and modes as directions in $T_{\Phi(w)}\mathfrak g$}
\label{sec:codim-form-modes}
Let's start defining:
\begin{equation}
  \label{eq:135}
  \Phi_w = \Phi-\Phi(0).
\end{equation}
This coincides with the $w$-dependent entries of the Higgs field.


5d and 7d modes  are infinitesimal deformations of the Higgs field $\Phi$, up to
gauge equivalence. It makes sense, if one is interested just in counting the number of linearly independent 7d elements supporting 5d localized modes, to identify them as tangent
directions
in $T_{\Phi}\mathfrak g$, with $\mathfrak{g}$ the 7d gauge algebra, transverse to the seven-dimensional gauge group orbits (since we already performed a gauge fixing). Indeed, we can interpret (\ref{codimension formula}) as a
statement on the tangent space $T_{\Phi(0)}\mathfrak g$: 5d modes are
directions
\begin{enumerate}
\item transverse to the nilpotent orbit $\mathcal O_0$,
\item tangent to the normal cone of $\mathfrak{g}$\footnote{In general, $\Phi(0) \in \mathcal O_0
    \hookrightarrow \text{Sing }(\mathcal N)$, and the tangent space is not
  well defined. The correct concept to use then is the one
of ``normal cone of $\mathcal N$ at the point $\Phi(0)$'':
$C_{\Phi(0)}\mathcal N$.}.
\end{enumerate}
Then, (\ref{codimension formula}) suggests that, if we find an isomorphism\footnote{$\mathcal
  N$ is, in general, a singular variety, but $\mathfrak g$ is everywhere smooth, and then $T_{\Phi(0)}\mathfrak g$ is well-defined.} (mentally
thinking $w \to 0$) 
\begin{equation}
  \label{eq:115}
  \Psi: T_{\Phi} \mathfrak g \to T_{\Phi(0)} \mathfrak g,
\end{equation}
then,
\begin{equation}
  \label{eq:116}
  \Psi(V_{5d})= T_{\Phi(0)}S_{0} \cap \textit{C}_{\Phi(0)} \mathcal N ,
\end{equation}
where $S_0$ is the Slodowy slice through $\Phi(0)$, $\textit{C}_{\Phi(0)}\mathcal N$ is the normal cone of $\mathcal N$ at the point $\Phi(0)$ and $V_{5d}$ is the linear subspace of $\mathfrak g$ supporting five-dimensional modes. We guess
(\ref{eq:116}) from (\ref{codimension formula}) because (\ref{codimension formula}) involves the
codimension of $\mathcal O_0$ inside the nilpotent cone, $S_0$ models
the transverse space to $\mathcal O_0$ at the point $\Phi(0) \in \mathcal
O_0$, and, to get something of dimension $n_{\text{ind}}$, we need to intersect
it with the correct ``ambient'' space (namely, the nilpotent cone and not the full algebra $\mathfrak g$). This
statement, at the level of tangent spaces, becomes (\ref{eq:116}).

  We used, to check if a mode  $\delta_{ij}$ belongs to
  $C_{\Phi(0)}\mathcal N$, the following condition
\begin{equation} \label{equation 117}
  \chi(\Phi+ \delta) = \chi(\Phi) + O(\delta^2).
\end{equation}
with $\chi$ indicating the characteristic polynomial.
\eqref{equation 117} defines the normal cone $C_{\Phi(0)}\mathcal N$, and generalizes the condition of ``tangency'' to points belonging to the singular locus of $\mathcal N$. We found by a case by case analysis that we can always perform the gauge-fixing in such a way that \eqref{equation 117} is respected for all the 5d modes in all the studied cases. On the other hand, we both checked in the analyzed cases and proven formally the transversality condition, that we deal with in the next section.

\subsubsection{Proof of the transversality of the 5d modes to $\mathcal
  O_0$}
\label{sec:proof-transv-5d}

Let us call $Y_0$ the nilnegative element of the standard triple
$\left\{H_0,Y_0, \Phi(0)\right\}$ in the sense of \cite{Collingwood}, with 
\begin{equation}
    H_{0} \equiv \comm{\Phi(0)}{Y_0},
\end{equation}
$\Phi(0)$ acts as a raising operator in
this triple. In the branching of $\mathfrak g$ under the $\left\{H_0,Y_0,
  \Phi(0)\right\}$, all the elements that are \textit{not the lowest weights states}
of their irreducible representation are in $\text{Im}(\text{ad}_{\Phi(0)})$ (the
reason is that there exists a lower weight state that was ``raised'' to them via
$\Phi(0)$), and can be completely gauged away. Viceversa, all the
elements that
are not in $\text{Im} (\text{ad}_{\Phi(0)})$ can not be \textit{completely} gauged away, and
produce either a  5d or a 7d mode. We can then say that
\begin{equation}
  \label{eq:119}
  \delta_{ij} \text{ is a (5d or 7d mode)} \Leftrightarrow \delta_{ij} \in \text{Ker}(Y_0),
\end{equation}
since $\text{Ker}(Y_0)$ defines the space of lowest weights states in the branching of $\mathfrak g$ under the triple.
Let's link (\ref{eq:119}) to the property of being transverse to $\mathcal
O_0$ at the point $\Phi(0)$. The transverse space to $\mathcal O_0$ in
$\mathfrak g$ is modeled by the Slodowy slice through $\Phi(0)$,
\begin{equation}
  \label{eq:120}
  S_0 \equiv \left\{Z \in \mathfrak g \mid \comm{Z- \Phi(0)}{Y_0}=0\right\}
  \textit{= } 
  \Phi(0) + \text{Ker}(Y_0),
\end{equation}
where $\Phi(0) + \text{Ker}(Y_0)$ is defined as the affine space through
$\Phi(0)$ in direction $\text{Ker}(Y_0)$, namely
\begin{equation}
  \label{eq:121}
  \Phi(0) + \text{Ker}(Y_0) \equiv \left\{Z \in \mathfrak g \mid Z =
    \Phi(0) + \lambda, \lambda \in \text{Ker}(Y_0)\right\}.
\end{equation}
The last equality in (\ref{eq:120}) means that (\ref{eq:119}) is equivalent say
that a necessary condition for an oscillation to be a (five or seven-dimensional) mode is to
being
along the transverse directions to $\mathcal O_0$ (namely, inside
$T_{\Phi(0)}S_0 < T_{\Phi(0)}\mathfrak g$).


\newpage
\section{Appendix C}\label{Appendix C}
\subsection{Explicit Higgs Branches for $\boldsymbol{(A_k,D_n)}$}

In the following pages we determine completely, as algebraic varieties, the Higgs Branches of M-theory on all the singularities $(A_k,D_n)$, with $k = 1, ..., 8$, and $n = 4,...,15$. There is nothing that forbids to continue the analysis for $k >8$ and $n > 15$, and our methods still apply, but we stopped here for space reasons. The tables are made up of four columns:
\begin{enumerate}
\item The first column indicates the Calabi-Yau threefold.
\item The second column indicates the $\text{Stab}(\Phi)$: the $U(1) $ factors form the flavor group, and the $\mathbb Z_{2}$ factors form the discrete gauging group. These groups were computed assuming simply-connected seven-dimensional gauge groups.
\item The third column contains matrices that describe how the five-dimensional modes localize w.r.t.\ the block decomposition of $\Phi$ into the blocks of table \ref{Table 2}. Each number corresponds to the amount of five-dimensional modes localized in that block. The colors represent the charges of the modes w.r.t\ the flavour and gauge groups, according to the key:
\begin{equation*}
\begin{cases}
\textcolor{black}{\text{black: }} \text{ uncharged modes} \\
\textcolor{red}{\text{red: }} \hspace{0.35cm}\text{ modes with charge }\pm 1\text{ under (possibly more than one) }U(1) \\
\quad\quad\text{ (and possibly one } \mathbb{Z}_2 \text{ factor}) \\
\textcolor{blue}{\text{blue: }} \hspace{0.2cm}\text{ modes with charge }\pm 2\text{ under }U(1) \text{\hspace{0.3cm}\ (and possibly one } \mathbb{Z}_2 \text{ factor}) \\
\textcolor{ForestGreen}{\text{green: }} \hspace{0.05cm}\text{ modes charged only under some } \mathbb{Z}_2 \text{ factors} \\
\end{cases}  
\end{equation*}
\item The last column indicates the quaternionic dimension of the Higgs branch, that coincides with the expected one \cite{Xie}. 
\end{enumerate}
Let us do an example of how to use the data in the tables to reconstruct the Higgs Branch. 
Let's pick, for example, the $(A_3,D_5)$ singularity (that is the case $k =2, n = 1$ of the family we have already examined in section \eqref{example Closset}). With our method, we find that  the stabilizers of the Higgs field are:
\begin{equation}
    \label{A3D5 stabilizers 2}
   \text{Stab}(\Phi) \equiv  \left(
\begin{array}{c|c|c|c}
 e^{i \alpha } &  0 & \mathbb 0 & \mathbb 0 \\
 \hline
  0 & e^{-i \alpha }& \mathbb 0 & \mathbb 0 \\
 \hline
 \mathbb 0 & \mathbb 0 & \epsilon _1 \mathbbm{1}_a & \mathbb 0 \\
 \hline
 \mathbb 0 & \mathbb 0 & \mathbb 0 & \epsilon _2 \mathbbm{1}_b\\
\end{array}
\right),
\end{equation}
with $\alpha \in \mathbb R, \epsilon_{1}, \epsilon_{2} = \pm 1$, $a = 4, b = 4$, and $\mathbb 0$ indicates the zero matrix of the appropriate size. The final result, with the data given in the tables, is independent from\footnote{Consequently in the table we do not give the data to compute $a,b$.} $a,b$ (but we needed them to compute the third column of the tables). In the second column of the tables below, we shortened \eqref{A3D5 stabilizers 2} as:
\begin{equation}
    \label{A3D5 stabilizers}
 \text{Stab}(\Phi) = \left(
\begin{array}{ccc}
 U(1) &  &  \\
  &\mathbb{Z}_2 &  \\
  &  & \mathbb{Z}_2\\
\end{array}
\right).  
\end{equation}
Notice that, passing from \eqref{A3D5 stabilizers 2} to \eqref{A3D5 stabilizers}, we have condensed the first two rows into a single row, and the first two columns into a single column: this is because the corresponding block of type (a) in the Higgs $\Phi$  (for the definition of the blocks of type (a) we refer to \nameref{Appendix A}) is a $2\times 2$ null matrix, that \textit{never} contains any localized mode.\\
\indent In more general cases, where the blocks of type (a) are not the null matrix, we explicitly keep them separated into two rows and two columns. In the $(A_3,D_4)$ case, for example, the stabilizer is written as:
\begin{equation}\label{stabilizer double}
   \text{Stab}(\Phi) =  \left(
\begin{array}{c|c|c|c}
 \textcolor{red}{e^{i \alpha }} &  0 & \mathbb 0 & \mathbb 0 \\
 \hline
  0 & \textcolor{red}{e^{-i \alpha }}& \mathbb 0 & \mathbb 0 \\
 \hline
 \mathbb 0 & \mathbb 0 & \textcolor{blue}{e^{i \beta }\mathbbm{1}_a} & \mathbb 0 \\
 \hline
 \mathbb 0 & \mathbb 0 & \mathbb 0 & \textcolor{blue}{e^{-i \beta } \mathbbm{1}_a}\\
\end{array}
\right)\equiv \left(
\begin{array}{ccc}
\textcolor{red}{ U(1)_{\alpha}} &  &  \\
  &\textcolor{blue}{U(1)_{\beta}} &  \\
  &  & \textcolor{blue}{U(1)_{\beta}} \\
\end{array}
\right),
\end{equation}
where $U(1)_{\alpha}$ refers to a vanishing $2\times 2$ block of type (a) in $\Phi$, and $U(1)_{\beta}$ refers to a non-vanishing block of type (a), that is therefore kept on two rows and two columns. That is the reason why $U(1)_{\beta}$ appears twice in \eqref{stabilizer double}.\\
\indent \textit{Summing up: we write stabilizers referred to blocks of type (a) on two rows and two columns, except when the block is a $2\times 2$ null matrix (in which case we write it only on one row and column).}\\

\indent The third column of the table indicates how the five-dimensional modes localize w.r.t.\ the block decomposition we highlighted in \eqref{A3D5 stabilizers 2}. In the case of the $(A_3,D_5)$, the five-dimensional modes distribute as follows:
\begin{equation}
    \label{A3D5 modes decomposition}
  \left(  \begin{array}{c|c|c|c}
  0  & 0 & \red{1} & \red{1}\\
 \hline
 0 & 0&  \red{1} & \red{1}\\
 \hline
 \red{1} &  \red{1} & 0 & \textcolor{ForestGreen}{4} \\
 \hline
 \red{1} &  \red{1} & \textcolor{ForestGreen}{4} & 0\\
\end{array}
\right),
\end{equation}
that translates in the table as: 
\begin{equation}\label{appendix C example}
\left(
\begin{array}{ccc}
 0 & \textcolor{red}{2} &\textcolor{red}{2} \\
 \textcolor{red}{2} & 0 &\textcolor{ForestGreen}{4} \\
 \textcolor{red}{2} & \textcolor{ForestGreen}{4} & 0 \\
\end{array}
\right),
\end{equation}
where as already explained above we have collapsed the first two rows and the first two columns, corresponding to a vanishing block of type (a) in $\Phi$, into a single one.\\

\newpage
\newgeometry{top=80pt,bottom=85pt,left=40pt,right=35pt}
\begin{table}[H]
  \centering
  \begin{tabular}{|L||P|P|L|}
    \hline
  \textbf{ CY} & \textbf{Stab} \boldsymbol{(\Phi)}  & \textbf{Modes}&  \boldsymbol{\textbf{dim}_{\mathbb H}HB}      \\
    \hline
    &&&
 \\[-15pt]
     (A_1,D_4)  &  \left(
\begin{array}{ccc}
 U(1)_{\alpha} &  &  \\
  & U(1)_{\beta} &  \\
  &  & U(1)_{\beta} \\
\end{array}
\right)   & \left(
\begin{array}{ccc}
 0 & \textcolor{red}{1} & \textcolor{red}{1} \\
 \textcolor{red}{1} & 0 & \textcolor{red}{1} \\
 \textcolor{red}{1} & \textcolor{red}{1} & 0 \\
\end{array}
    \right)  & 3     \\           
   &&& \\[-15pt]
    \hline
   &&& \\[-15pt]
     (A_1,D_5)  &  \left(
\begin{array}{cc}
 U(1) &  \\
  & \mathbb Z_2 \\
\end{array}
\right)      &   \left(
\begin{array}{cc}
 0 & \textcolor{red}{2} \\
 \textcolor{red}{2} & 2 \\
\end{array}
    \right)
     & 3                \\
     &&& \\[-15pt]
    \hline
   &&& \\[-15pt]
     (A_1,D_6)  &  \left(
\begin{array}{ccc}
 U(1)_{\alpha} &  &  \\
  & U(1)_{\beta} &  \\
  &  & U(1)_{\beta} \\
\end{array}
\right)   & \left(
\begin{array}{ccc}
 0 & \textcolor{red}{1} & \textcolor{red}{1} \\
 \textcolor{red}{1} & 0 & \textcolor{blue}{2} \\
 \textcolor{red}{1} & \textcolor{blue}{2} & 0 \\
\end{array}
    \right)  & 4                \\
     &&& \\[-15pt]
    \hline
   &&& \\[-15pt]
         (A_1,D_7)  &  \left(
\begin{array}{cc}
 U(1) &  \\
  & \mathbb Z_2 \\
\end{array}
\right)      &   \left(
\begin{array}{cc}
 0 & \textcolor{red}{2} \\
 \textcolor{red}{2} & 4 \\
\end{array}
    \right)
     & 4                \\
      &&& \\[-15pt]
    \hline
   &&& \\[-15pt]
             (A_1,D_8)  &  \left(
\begin{array}{ccc}
 U(1)_{\alpha} &  &  \\
  & U(1)_{\beta} &  \\
  &  & U(1)_{\beta} \\
\end{array}
\right)   & \left(
\begin{array}{ccc}
 0 & \textcolor{red}{1} & \textcolor{red}{1} \\
 \textcolor{red}{1} & 0 & \textcolor{blue}{3} \\
 \textcolor{red}{1} & \textcolor{blue}{3} & 0 \\
\end{array}
    \right)  & 5                \\
     &&& \\[-15pt]
    \hline
   &&& \\[-15pt]
         (A_1,D_9)  &  \left(
\begin{array}{cc}
 U(1) &  \\
  & \mathbb Z_2 \\
\end{array}
\right)      &   \left(
\begin{array}{cc}
 0 & \textcolor{red}{2} \\
 \textcolor{red}{2} & 6 \\
\end{array}
    \right)
     & 5                \\[30pt]
    \hline

  \end{tabular}
\end{table}

\newpage
\begin{table}[H]
  \centering
  \begin{tabular}{|L||P|P|L|}
    \hline
 \textbf{ CY} & \textbf{Stab} \boldsymbol{(\Phi)}  & \textbf{Modes}&  \boldsymbol{\textbf{dim}_{\mathbb H}HB}      \\
    \hline
    &&&
 \\[-15pt]
   (A_1,D_{10})  &  \left(
\begin{array}{ccc}
 U(1)_{\alpha} &  &  \\
  & U(1)_{\beta} &  \\
  &  & U(1)_{\beta} \\
\end{array}
\right)   & \left(
\begin{array}{ccc}
 0 & \textcolor{red}{1} & \textcolor{red}{1} \\
 \textcolor{red}{1} & 0 & \textcolor{blue}{4} \\
 \textcolor{red}{1} & \textcolor{blue}{4} & 0 \\
\end{array}
    \right)  & 6                \\
     &&& \\[-15pt]
    \hline
   &&& \\[-15pt]
   (A_1,D_{11})  &  \left(
\begin{array}{cc}
 U(1) &  \\
  & \mathbb Z_2 \\
\end{array}
\right)      &   \left(
\begin{array}{cc}
 0 &  \textcolor{red}{2} \\
  \textcolor{red}{2} & 8 \\
\end{array}
    \right)
     & 6                \\
     &&& \\[-15pt]
    \hline
   &&& \\[-15pt]
            
 (A_1,D_{12})  &  \left(
\begin{array}{ccc}
 U(1)_{\alpha} &  &  \\
  & U(1)_{\beta} &  \\
  &  & U(1)_{\beta} \\
\end{array}
\right)   & \left(
\begin{array}{ccc}
 0 &  \textcolor{red}{1} & \textcolor{red}{1} \\
  \textcolor{red}{1} & 0 & \textcolor{blue}{5} \\
  \textcolor{red}{1} & \textcolor{blue}{5} & 0 \\
\end{array}
    \right)  & 7                \\
     &&& \\[-15pt]
    \hline
   &&& \\[-15pt]
         (A_1,D_{13})  &  \left(
\begin{array}{cc}
 U(1) &  \\
  & \mathbb Z_2 \\
\end{array}
\right)      &   \left(
\begin{array}{cc}
 0 & \textcolor{red}{2} \\
 \textcolor{red}{2} & 10 \\
\end{array}
    \right)
     & 7                \\
     &&& \\[-15pt]
    \hline
   &&& \\[-15pt]
             (A_1,D_{14})  &  \left(
\begin{array}{ccc}
 U(1)_{\alpha} &  &  \\
  & U(1)_{\beta} &  \\
  &  & U(1)_{\beta} \\
\end{array}
\right)   & \left(
\begin{array}{ccc}
 0 & \textcolor{red}{1} & \textcolor{red}{1} \\
 \textcolor{red}{1} & 0 & \textcolor{blue}{6} \\
 \textcolor{red}{1} & \textcolor{blue}{6} & 0 \\
\end{array}
    \right)  & 8                \\
     &&& \\[-15pt]
    \hline
   &&& \\[-15pt]
             (A_1,D_{15})  &  \left(
\begin{array}{cc}
 U(1) &  \\
  & \mathbb Z_2 \\
\end{array}
\right)      &   \left(
\begin{array}{cc}
 0 & \textcolor{red}{2} \\
 \textcolor{red}{2} & 12 \\
\end{array}
    \right)  & 8                \\[30pt]
 \hline
  \end{tabular}
\end{table}

\begin{table}[H]
  \centering
  \begin{tabular}{|L||P|P|L|}
    \hline
 \textbf{ CY} & \textbf{Stab} \boldsymbol{(\Phi)}  & \textbf{Modes}&  \boldsymbol{\textbf{dim}_{\mathbb H}HB}      \\
    \hline
    &&&
 \\[-15pt]
     (A_2,D_4)  &  \left(
\begin{array}{cc}
  \mathbb{Z}_2 &  \\
 &  \mathbb{Z}_2 \\
\end{array}
\right)  & \left(
\begin{array}{cc}
 0 &\textcolor{ForestGreen}{4} \\
 \textcolor{ForestGreen}{4} & 0 \\
\end{array}
\right)  & 4                \\
 &&& \\[-15pt]
    \hline
   &&& \\[-15pt]
     (A_2,D_5)  &  \left(
\begin{array}{c}
\mathbb Z_2 \\
\end{array}
\right)      &   \left(
\begin{array}{c}
 10 
\end{array}
    \right)
     & 5                \\
     &&& \\[-15pt]
    \hline
   &&& \\[-15pt]
     (A_2,D_6)  &  \left(
\begin{array}{c}
\mathbb Z_2 \\
\end{array}
\right)      &   \left(
\begin{array}{c}
 12 
\end{array}
    \right)
     & 6                \\
     &&& \\[-15pt]
    \hline
   &&& \\[-15pt]
         (A_2,D_7)  &  \left(
\begin{array}{cc}
 \mathbb Z_2&  \\
  & \mathbb Z_2 \\
\end{array}
\right)      &   \left(
\begin{array}{cc}
 0 & \textcolor{ForestGreen}{6} \\
\textcolor{ForestGreen}{6} & 2 \\
\end{array}
    \right)
     & 7                \\
     &&& \\[-15pt]
    \hline
   &&& \\[-15pt]
             (A_2,D_8)  &  \left(
\begin{array}{c}
\mathbb Z_2 \\
\end{array}
\right)      &   \left(
\begin{array}{c}
 16 
\end{array}
    \right)
     & 8                \\
     &&& \\[-15pt]
    \hline
   &&& \\[-15pt]
         (A_2,D_9)  &  \left(
\begin{array}{c}
\mathbb Z_2 \\
\end{array}
\right)      &   \left(
\begin{array}{c}
 18 
\end{array}
    \right)
     & 9                \\
     &&& \\[-15pt]
    \hline
   &&& \\[-15pt]
             (A_2,D_{10})  &  \left(
\begin{array}{cc}
  \mathbb{Z}_2 &  \\
 &  \mathbb{Z}_2 \\
\end{array}
\right)  & \left(
\begin{array}{cc}
 0 & \textcolor{ForestGreen}{8} \\
 \textcolor{ForestGreen}{8} & 4 \\
\end{array}
\right)  & 10                \\
 &&& \\[-15pt]
    \hline
   &&& \\[-15pt]
         (A_2,D_{11})  &  \left(
\begin{array}{c}
\mathbb Z_2 \\
\end{array}
\right)      &   \left(
\begin{array}{c}
 22 
\end{array}
    \right)
     &11                \\
     &&& \\[-15pt]
    \hline
   &&& \\[-15pt]
             (A_2,D_{12})  &  \left(
\begin{array}{c}
\mathbb Z_2 \\
\end{array}
\right)      &   \left(
\begin{array}{c}
 24 
\end{array}
    \right)
  & 12                \\
  &&& \\[-15pt]
    \hline
   &&& \\[-15pt]
         (A_2,D_{13})  &  \left(
\begin{array}{cc}
\mathbb Z_2 &  \\
  & \mathbb Z_2 \\
\end{array}
\right)      &   \left(
\begin{array}{cc}
 0 & \textcolor{ForestGreen}{10} \\
 \textcolor{ForestGreen}{10} & 6 \\
\end{array}
    \right)
     & 13                \\[20pt]

        \hline
  \end{tabular}
\end{table}

\newpage
\begin{table}[H]
  \centering
  \begin{tabular}{|L||P|P|L|}
    \hline
  \textbf{ CY} & \textbf{Stab} \boldsymbol{(\Phi)}  & \textbf{Modes}&  \boldsymbol{\textbf{dim}_{\mathbb H}HB}      \\
    \hline
    &&&
 \\[-15pt]
             (A_2,D_{14})  &  \left(
\begin{array}{c}
\mathbb Z_2 \\
\end{array}
\right)      &   \left(
\begin{array}{c}
 28 
\end{array}
    \right)
 & 14                \\
 &&& \\[-15pt]
    \hline
   &&& \\[-15pt]
             (A_2,D_{15})  &  \left(
\begin{array}{c}
\mathbb Z_2 \\
\end{array}
\right)      &   \left(
\begin{array}{c}
30 
\end{array}
    \right)  & 15                \\[20pt]
        
        \hline
  \end{tabular}
\end{table}

\begin{table}[H]
  \centering
  \begin{tabular}{|L||P|P|L|}
    \hline
 \textbf{ CY} & \textbf{Stab} \boldsymbol{(\Phi)}  & \textbf{Modes}&  \boldsymbol{\textbf{dim}_{\mathbb H}HB}      \\
    \hline
    &&&
 \\[-15pt]
     (A_3,D_{4})  &  \scalemath{0.9}{\left(
\begin{array}{ccc}
 U(1)_{\alpha} &  &  \\
  &U(1)_{\beta} &  \\
  &  & U(1)_{\beta} \\
\end{array}
\right)}      &   \left(
\begin{array}{ccc}
 0 & \textcolor{red}{2} & \textcolor{red}{2} \\
 \textcolor{red}{2} & 1 & \textcolor{blue}{2} \\
 \textcolor{red}{2} & \textcolor{blue}{2} & 1 \\
\end{array}
\right)
 & 7                \\
&&& \\[-15pt]
    \hline
   &&& \\[-15pt]
     (A_3,D_5)  &  \left(
\begin{array}{ccc}
 U(1) &  &  \\
  &\mathbb{Z}_2 &  \\
  &  & \mathbb{Z}_2\\
\end{array}
\right)      &   \left(
\begin{array}{ccc}
 0 & \textcolor{red}{2} &\textcolor{red}{2} \\
 \textcolor{red}{2} & 0 &\textcolor{ForestGreen}{4} \\
 \textcolor{red}{2} & \textcolor{ForestGreen}{4} & 0 \\
\end{array}
\right)
 & 8                \\
&&& \\[-15pt]
    \hline
   &&& \\[-15pt]
     (A_3,D_{6})  &  \left(
\begin{array}{ccc}
 U(1)_{\alpha} &  &  \\
  &U(1)_{\beta} &  \\
  &  & U(1)_{\beta} \\
\end{array}
\right)      &   \left(
\begin{array}{ccc}
 0 & \textcolor{red}{2} & \textcolor{red}{2} \\
 \textcolor{red}{2} & 2 & \textcolor{blue}{4}  \\
 \textcolor{red}{2} & \textcolor{blue}{4} & 2 \\
\end{array}
\right)
 & 10                      \\[45pt]

   \hline
  \end{tabular}
\end{table} 

 \newpage
\begin{table}[H]
  \centering
  \begin{tabular}{|L||P|P|L|}
    \hline
   \textbf{ CY} & \textbf{Stab} \boldsymbol{(\Phi)}  & \textbf{Modes}&  \boldsymbol{\textbf{dim}_{\mathbb H}HB}      \\
    \hline
    &&&
 \\[-15pt]

    (A_3,D_{7})  &  \scalemath{0.7}{\left(
\begin{array}{ccccc}
 U(1)_{\alpha} &  &  &  &  \\
  &  U(1)_{\beta}  &  &  &  \\
  &  &  U(1)_{\beta}  &  &  \\
  &  &  &  U(1)_{\gamma}  &  \\
  &  &  &  & U(1)_{\gamma}   \\
\end{array}
\right)}      &   \scalemath{0.7}{\left(
\begin{array}{ccccc}
 0 & \textcolor{red}{1} & \textcolor{red}{1}  & \textcolor{red}{1}  & \textcolor{red}{1}  \\
 \textcolor{red}{1}  & 0 & \textcolor{blue}{1}  & \textcolor{red}{3}  & \textcolor{red}{3}  \\
 \textcolor{red}{1}  & \textcolor{blue}{1}  & 0 & \textcolor{red}{3} & \textcolor{red}{3}  \\
 \textcolor{red}{1}  & 0 & 0 & 0 & \textcolor{blue}{1}  \\
 \textcolor{red}{1}  & 0 & 0 & \textcolor{blue}{1}  & 0 \\
\end{array}
\right)}
 & 12                \\
&&& \\[-15pt]
    \hline
   &&& \\[-15pt]
    (A_3,D_{8})  &  \left(
\begin{array}{ccc}
 U(1)_{\alpha} &  &  \\
  &U(1)_{\beta} &  \\
  &  & U(1)_{\beta} \\
\end{array}
\right)      &  \left(
\begin{array}{ccc}
 0 & \textcolor{red}{2} & \textcolor{red}{2}  \\
 \textcolor{red}{2}  & 3 & \textcolor{blue}{6}  \\
 \textcolor{red}{2}  & \textcolor{blue}{6} & 3 \\
\end{array}
\right)
 & 13         \\
&&& \\[-15pt]
    \hline
   &&& \\[-15pt]
    (A_3,D_{9})  &  \left(
\begin{array}{ccc}
 U(1) &  &  \\
  &\mathbb{Z}_2 &  \\
  &  & \mathbb{Z}_2\\
\end{array}
\right)      &  \left(
\begin{array}{ccc}
 0 & \textcolor{red}{2} & \textcolor{red}{2} \\
 \textcolor{red}{2} & 2 & \textcolor{ForestGreen}{8} \\
 \textcolor{red}{2} & \textcolor{ForestGreen}{8} & 2 \\
\end{array}
\right)
 & 14                \\
&&& \\[-15pt]
    \hline
   &&& \\[-15pt]
    (A_3,D_{10})  &  \left(
\begin{array}{ccc}
 U(1)_{\alpha} &  &  \\
  &U(1)_{\beta} &  \\
  &  & U(1)_{\beta} \\
\end{array}
\right)  &  \left(
\begin{array}{ccc}
 0 & \textcolor{red}{2} & \textcolor{red}{2} \\
 \textcolor{red}{2} & 4 & \textcolor{blue}{8} \\
 \textcolor{red}{2} & \textcolor{blue}{8} & 4 \\
\end{array}
\right)
 & 16                \\
&&& \\[-15pt]
    \hline
   &&& \\[-15pt]
    (A_3,D_{11})  &  \scalemath{0.7}{\left(
\begin{array}{ccccc}
 U(1)_{\alpha} &  &  &  &  \\
  &  U(1)_{\beta}  &  &  &  \\
  &  &  U(1)_{\beta}  &  &  \\
  &  &  &  U(1)_{\gamma}  &  \\
  &  &  &  & U(1)_{\gamma}   \\
\end{array}
\right)}      &  \scalemath{0.7}{\left(
\begin{array}{ccccc}
 0 & \textcolor{red}{1} & \textcolor{red}{1} & \textcolor{red}{1} & \textcolor{red}{1} \\
 \textcolor{red}{1} & 0 & \textcolor{blue}{2} & \textcolor{red}{5} & \textcolor{red}{5} \\
 \textcolor{red}{1} & \textcolor{blue}{2} & 0 & \textcolor{red}{5} & \textcolor{red}{5} \\
 \textcolor{red}{1} & 0 & 0 & 0 & \textcolor{blue}{2} \\
 \textcolor{red}{1} & 0 & 0 & \textcolor{blue}{2} & 0 \\
\end{array}
\right)}  
& 18                  
             \\[50pt] 
   \hline
  \end{tabular}
\end{table}

    \newpage
\begin{table}[H]
  \centering
  \begin{tabular}{|L||P|P|L|}
    \hline
   \textbf{ CY} & \textbf{Stab} \boldsymbol{(\Phi)}  & \textbf{Modes}&  \boldsymbol{\textbf{dim}_{\mathbb H}HB}      \\
    \hline
    &&&
 \\[-15pt]

    (A_3,D_{12})  &  \left(
\begin{array}{ccc}
 U(1)_{\alpha} &  &  \\
  &U(1)_{\beta} &  \\
  &  & U(1)_{\beta} \\
\end{array}
\right)      &  \left(
\begin{array}{ccc}
 0 & \textcolor{red}{2} & \textcolor{red}{2} \\
 \textcolor{red}{2} & 5 & \textcolor{blue}{10} \\
 \textcolor{red}{2} & \textcolor{blue}{10} & 5 \\
\end{array}
\right)
 & 19                \\
&&& \\[-15pt]
    \hline
   &&& \\[-15pt]
    (A_3,D_{13})  &  \left(
\begin{array}{ccc}
 U(1) &  &  \\
  &\mathbb{Z}_2 &  \\
  &  & \mathbb{Z}_2\\
\end{array}
\right)      &   \left(
\begin{array}{ccc}
 0 & \textcolor{red}{2} & \textcolor{red}{2} \\
 \textcolor{red}{2} & 4 & \textcolor{ForestGreen}{12} \\
\textcolor{red}{2} & \textcolor{ForestGreen}{12} & 4 \\
\end{array}
\right)
 & 20         \\
&&& \\[-15pt]
    \hline
   &&& \\[-15pt]      
    (A_3,D_{14})  &  \left(
\begin{array}{ccc}
 U(1)_{\alpha} &  &  \\
  &U(1)_{\beta} &  \\
  &  & U(1)_{\beta} \\
\end{array}
\right)      &   \left(
\begin{array}{ccc}
 0 & \textcolor{red}{2} & \textcolor{red}{2} \\
 \textcolor{red}{2} & 6 & \textcolor{blue}{12} \\
 \textcolor{red}{2} & \textcolor{blue}{12} & 6 \\
\end{array}
\right)
 & 22   \\
&&& \\[-15pt]
    \hline
   &&& \\[-15pt]
       (A_3,D_{15})  &  \scalemath{0.7}{\left(
\begin{array}{ccccc}
 U(1)_{\alpha} &  &  &  &  \\
  &  U(1)_{\beta}  &  &  &  \\
  &  &  U(1)_{\beta}  &  &  \\
  &  &  &  U(1)_{\gamma}  &  \\
  &  &  &  & U(1)_{\gamma}   \\
\end{array}
\right)}      &  \scalemath{0.7}{\left(
\begin{array}{ccccc}
 0 & \textcolor{red}{1} & \textcolor{red}{1} & \textcolor{red}{1} & \textcolor{red}{1} \\
 \textcolor{red}{1} & 0 & \textcolor{blue}{3}  & \textcolor{red}{7} & \textcolor{red}{7}  \\
 \textcolor{red}{1} & \textcolor{blue}{3}  & 0 & \textcolor{red}{7}  & \textcolor{red}{7}  \\
 \textcolor{red}{1} & 0 & 0 & 0 & \textcolor{blue}{3}  \\
 \textcolor{red}{1} & 0 & 0 & \textcolor{blue}{3}  & 0 \\
\end{array}
\right)} 
 & 24                \\[50pt] 
   \hline
  \end{tabular}
\end{table}

\newpage

\begin{table}[H]
  \centering
  \begin{tabular}{|L||P|P|L|}
    \hline
  \textbf{ CY} & \textbf{Stab} \boldsymbol{(\Phi)}  & \textbf{Modes}&  \boldsymbol{\textbf{dim}_{\mathbb H}HB}      \\
    \hline
    &&&
 \\[-15pt]
             (A_4,D_{4})  &  \left(
\begin{array}{c}
 \mathbb{Z}_2 \\
\end{array}
\right)      &   \left(
\begin{array}{c}
 16 
\end{array}
    \right)
 & 8                \\
 &&& \\[-15pt]
    \hline
   &&& \\[-15pt]
  (A_4,D_{5})  &  \left(
\begin{array}{c}
 \mathbb{Z}_2 \\
\end{array}
\right)      &   \left(
\begin{array}{c}
 20 
\end{array}
    \right)
 & 10                \\
 &&& \\[-15pt]
    \hline
   &&& \\[-15pt]
 (A_4,D_{6})  &  \left(
\begin{array}{ccc}
 \mathbb{Z}_2 &  &  \\
  & \mathbb{Z}_2 &  \\
  &  & \mathbb{Z}_2 \\
\end{array}
\right)      &  \left(
\begin{array}{ccc}
 0 & \textcolor{ForestGreen}{4} & \textcolor{ForestGreen}{4} \\
 \textcolor{ForestGreen}{4} & 0 & \textcolor{ForestGreen}{4} \\
 \textcolor{ForestGreen}{4} & \textcolor{ForestGreen}{4} & 0 \\
\end{array}
\right)
 & 12                \\
 &&& \\[-15pt]
    \hline
   &&& \\[-15pt]
  (A_4,D_{7})  &  \left(
\begin{array}{c}
 \mathbb{Z}_2 \\
\end{array}
\right)      &   \left(
\begin{array}{c}
 28 
\end{array}
    \right)
 & 14                \\
 &&& \\[-15pt]
    \hline
   &&& \\[-15pt]
     (A_4,D_{8})  &  \left(
\begin{array}{c}
 \mathbb{Z}_2 \\
\end{array}
\right)      &   \left(
\begin{array}{c}
 32 
\end{array}
    \right)
 & 16                \\
 &&& \\[-15pt]
    \hline
   &&& \\[-15pt]
     (A_4,D_{9})  &  \left(
\begin{array}{c}
 \mathbb{Z}_2 \\
\end{array}
\right)      &   \left(
\begin{array}{c}
 36
\end{array}
    \right)
 & 18                \\
 &&& \\[-15pt]
    \hline
   &&& \\[-15pt]
     (A_4,D_{10})  &  \left(
\begin{array}{c}
 \mathbb{Z}_2 \\
\end{array}
\right)      &   \left(
\begin{array}{c}
 40 
\end{array}
    \right)
 & 20                \\
 &&& \\[-15pt]
    \hline
   &&& \\[-15pt]
 (A_4,D_{11})  &  \left(
\begin{array}{ccc}
 \mathbb{Z}_2 &  &  \\
  & \mathbb{Z}_2 &  \\
  &  & \mathbb{Z}_2 \\
\end{array}
\right)      &  \left(
\begin{array}{ccc}
 0 & \textcolor{ForestGreen}{6} & \textcolor{ForestGreen}{6} \\
 \textcolor{ForestGreen}{6} & 2 & \textcolor{ForestGreen}{8} \\
 \textcolor{ForestGreen}{6} & \textcolor{ForestGreen}{8} & 2 \\
\end{array}
\right)
 & 22               \\
 &&& \\[-15pt]
    \hline
   &&& \\[-15pt]
   (A_4,D_{12})  &  \left(
\begin{array}{c}
 \mathbb{Z}_2 \\
\end{array}
\right)      &   \left(
\begin{array}{c}
 48 
\end{array}
    \right)
 & 24                \\[20pt]
   \hline
  \end{tabular}
\end{table}

\newpage
\begin{table}[H]
  \centering
  \begin{tabular}{|L||P|P|L|}
    \hline
 \textbf{ CY} & \textbf{Stab} \boldsymbol{(\Phi)}  & \textbf{Modes}&  \boldsymbol{\textbf{dim}_{\mathbb H}HB}      \\
    \hline
    &&&
 \\[-15pt]
      (A_4,D_{13})  &  \left(
\begin{array}{c}
 \mathbb{Z}_2 \\
\end{array}
\right)      &   \left(
\begin{array}{c}
 52 
\end{array}
    \right)
 & 26                \\
 &&& \\[-15pt]
    \hline
   &&& \\[-15pt]
      (A_4,D_{14})  &  \left(
\begin{array}{c}
 \mathbb{Z}_2 \\
\end{array}
\right)      &   \left(
\begin{array}{c}
 56 
\end{array}
    \right)
 & 28                \\
 &&& \\[-15pt]
    \hline
   &&& \\[-15pt]
      (A_4,D_{15})  &  \left(
\begin{array}{c}
 \mathbb{Z}_2 \\
\end{array}
\right)      &   \left(
\begin{array}{c}
 60 
\end{array}
    \right)
 & 30                \\[25pt]

   \hline
  \end{tabular}
\end{table}

\begin{table}[H]
  \centering
  \begin{tabular}{|L||P|P|L|}
    \hline
 \textbf{ CY} & \textbf{Stab} \boldsymbol{(\Phi)}  & \textbf{Modes}&  \boldsymbol{\textbf{dim}_{\mathbb H}HB}      \\
    \hline
    &&&
 \\[-15pt]
   (A_5,D_{4})  &  \left(
\begin{array}{cccc}
  U(1)_{\alpha}&  &  &  \\
   &U(1)_{\beta}  &  &  \\
     &  & U(1)_{\beta} &  \\
       &  &  & \mathbb{Z}_2 \\
\end{array}
\right)      &   \left(
\begin{array}{cccc}
 0 & \textcolor{red}{1} & \textcolor{red}{1} &  \textcolor{red}{4} \\
 \textcolor{red}{1} & 0 & 0 & \textcolor{red}{2} \\
 \textcolor{red}{1} & 0 & 0 &  \textcolor{red}{2}\\
  \textcolor{red}{4} & \textcolor{red}{2} &   \textcolor{red}{2} & 2 \\
\end{array}
\right)
 & 11                \\
&&& \\[-15pt]
    \hline
   &&& \\[-15pt]
     (A_5,D_{5})  &  \left(
\begin{array}{cc}
 U(1) &  \\
  & \mathbb{Z}_2 \\
\end{array}
\right)      &   \left(
\begin{array}{cc}
 0 & \textcolor{red}{6} \\
 \textcolor{red}{6} & 14 \\
\end{array}
\right)
 & 13                \\
&&& \\[-15pt]
    \hline
   &&& \\[-15pt]
     (A_5,D_{6})  &  \left(
\begin{array}{ccc}
 U(1)_{\alpha} &  &  \\
  & U(1)_{\beta}  &  \\
  &  &  U(1)_{\beta} \\
\end{array}
\right)      &   \left(
\begin{array}{ccc}
 0 & \textcolor{red}{3} & \textcolor{red}{3} \\
 \textcolor{red}{3} & 4 & \textcolor{blue}{6} \\
 \textcolor{red}{3} & \textcolor{blue}{6} & 4 \\
\end{array}
\right)
 & 16                 \\[50pt]

   \hline
  \end{tabular}
\end{table}

\newpage
\begin{table}[H]
  \centering
  \begin{tabular}{|L||P|P|L|}
    \hline
 \textbf{ CY} & \textbf{Stab} \boldsymbol{(\Phi)}  & \textbf{Modes}&  \boldsymbol{\textbf{dim}_{\mathbb H}HB}      \\
    \hline
    &&&
 \\[-15pt]
  
    (A_5,D_{7})  &  \left(
\begin{array}{cccc}
  U(1)_{\alpha}&  &  &  \\
   &\mathbb{Z}_2  &  &  \\
     &  & \mathbb{Z}_2 &  \\
       &  &  & \mathbb{Z}_2 \\
\end{array}
\right)      &   \left(
\begin{array}{cccc}
 0 & \textcolor{red}{2} & \textcolor{red}{2} & \textcolor{red}{2} \\
 \textcolor{red}{2} & 0 & \textcolor{ForestGreen}{4} & \textcolor{ForestGreen}{4}  \\
 \textcolor{red}{2} & \textcolor{ForestGreen}{4}  & 0 & \textcolor{ForestGreen}{4}  \\
 \textcolor{red}{2} & \textcolor{ForestGreen}{4}  & \textcolor{ForestGreen}{4}  & 0 \\
\end{array}
\right)
 & 18                \\
&&& \\[-15pt]
    \hline
   &&& \\[-15pt]
    (A_5,D_{8})  &  \left(
\begin{array}{ccc}
 U(1)_{\alpha} &  &  \\
  & U(1)_{\beta}  &  \\
  &  &  U(1)_{\beta} \\
\end{array}
\right)      &   \left(
\begin{array}{ccc}
 0 & \textcolor{red}{3} & \textcolor{red}{3} \\
\textcolor{red}{3} & 6 & \textcolor{blue}{9} \\
\textcolor{red}{3} & \textcolor{blue}{9}  & 6 \\
\end{array}
\right)
 & 21                \\
&&& \\[-15pt]
    \hline
   &&& \\[-15pt]
    (A_5,D_{9})  &  \left(
\begin{array}{cc}
 U(1) &  \\
  & \mathbb{Z}_2 \\
\end{array}
\right)      &   \left(
\begin{array}{cc}
 0 & \textcolor{red}{6} \\
 \textcolor{red}{6} & 34 \\
\end{array}
\right)
 & 23                \\
&&& \\[-15pt]
    \hline
   &&& \\[-15pt]
    (A_5,D_{10})  &  \left(
\begin{array}{cccc}
  U(1)_{\alpha}&  &  &  \\
   &U(1)_{\beta}  &  &  \\
     &  & U(1)_{\beta} &  \\
       &  &  & \mathbb{Z}_2 \\
\end{array}
\right)      &   \left(
\begin{array}{cccc}
 0 & \textcolor{red}{1} & \textcolor{red}{1} & \textcolor{red}{4} \\
 \textcolor{red}{1} & 0 & \textcolor{blue}{1} & \textcolor{red}{6} \\
 \textcolor{red}{1} & \textcolor{blue}{1}  & 0 & \textcolor{red}{6}  \\
 \textcolor{red}{4} & \textcolor{red}{6}  & \textcolor{red}{6}  & 14 \\
\end{array}
\right)
 & 26                \\
&&& \\[-15pt]
    \hline
   &&& \\[-15pt]
    (A_5,D_{11})  &  \left(
\begin{array}{cc}
 U(1) &  \\
  & \mathbb{Z}_2 \\
\end{array}
\right)      &   \left(
\begin{array}{cc}
 0 & \textcolor{red}{6} \\
 \textcolor{red}{6} & 44 \\
\end{array}
\right)
 & 28                \\
&&& \\[-15pt]
    \hline
   &&& \\[-15pt]
    (A_5,D_{12})  &  \left(
\begin{array}{ccc}
 U(1)_{\alpha} &  &  \\
  & U(1)_{\beta}  &  \\
  &  &  U(1)_{\beta} \\
\end{array}
\right)      &  \left(
\begin{array}{ccc}
 0 & \textcolor{red}{3} & \textcolor{red}{3} \\
 \textcolor{red}{3} & 10 & \textcolor{blue}{15} \\
 \textcolor{red}{3} & \textcolor{blue}{15} & 10 \\
\end{array}
\right)
 & 31                        \\[50pt]

   \hline
  \end{tabular}
\end{table}

\newpage
\begin{table}[H]
  \centering
  \begin{tabular}{|L||P|P|L|}
    \hline
 \textbf{ CY} & \textbf{Stab} \boldsymbol{(\Phi)}  & \textbf{Modes}&  \boldsymbol{\textbf{dim}_{\mathbb H}HB}      \\
    \hline
    &&&
 \\[-15pt]
    (A_5,D_{13})  &  \left(
\begin{array}{cccc}
  U(1)&  &  &  \\
   &\mathbb{Z}_2  &  &  \\
     &  & \mathbb{Z}_2 &  \\
       &  &  & \mathbb{Z}_2 \\
\end{array}
\right)      &   \left(
\begin{array}{cccc}
 0 & \textcolor{red}{2} &\textcolor{red}{2} & \textcolor{red}{2} \\
 \textcolor{red}{2} & 2 & \textcolor{ForestGreen}{8} &  \textcolor{ForestGreen}{8} \\
 \textcolor{red}{2} &  \textcolor{ForestGreen}{8} & 2 &  \textcolor{ForestGreen}{8} \\
 \textcolor{red}{2} &  \textcolor{ForestGreen}{8} &  \textcolor{ForestGreen}{8} & 2 \\
\end{array}
\right)
 & 33               \\
&&& \\[-15pt]
    \hline
   &&& \\[-15pt]
    (A_5,D_{14})  &  \left(
\begin{array}{ccc}
 U(1)_{\alpha} &  &  \\
  & U(1)_{\beta}  &  \\
  &  &  U(1)_{\beta} \\
\end{array}
\right)      &   \left(
\begin{array}{ccc}
 0 & \textcolor{red}{3} & \textcolor{red}{3} \\
 \textcolor{red}{3} & 12 & \textcolor{blue}{18} \\
 \textcolor{red}{3} & \textcolor{blue}{18} & 12 \\
\end{array}
\right)
 & 36                \\
&&& \\[-15pt]
    \hline
   &&& \\[-15pt]
       (A_5,D_{15})  &  \left(
\begin{array}{cc}
 U(1) &  \\
  & \mathbb{Z}_2 \\
\end{array}
\right)      &   \left(
\begin{array}{cc}
 0 & \textcolor{red}{6}\\
 \textcolor{red}{6} & 64 \\
\end{array}
\right)
 & 38        \\[30pt]

   \hline
  \end{tabular}
\end{table} 

\begin{table}[H]
  \centering
  \begin{tabular}{|L||P|P|L|}
    \hline
 \textbf{ CY} & \textbf{Stab} \boldsymbol{(\Phi)}  & \textbf{Modes}&  \boldsymbol{\textbf{dim}_{\mathbb H}HB}      \\
    \hline
    &&&
 \\[-15pt]
    
     (A_6,D_{4})  &  \left(
\begin{array}{c}
 \mathbb{Z}_2 \\
\end{array}
\right)      &   \left(
\begin{array}{c}
 24 
\end{array}
    \right)
 & 12                \\
 &&& \\[-15pt]
    \hline
   &&& \\[-15pt]
      (A_6,D_{5})  &  \left(
\begin{array}{c}
 \mathbb{Z}_2 \\
\end{array}
\right)      &   \left(
\begin{array}{c}
 30 
\end{array}
    \right)
 & 15                \\
 &&& \\[-15pt]
    \hline
   &&& \\[-15pt]
      (A_6,D_{6})  &  \left(
\begin{array}{c}
 \mathbb{Z}_2 \\
\end{array}
\right)      &   \left(
\begin{array}{c}
 36 
\end{array}
    \right)
 & 18                \\
 &&& \\[-15pt]
    \hline
   &&& \\[-15pt]
   (A_6,D_{7})  &  \left(
\begin{array}{c}
 \mathbb{Z}_2 \\
\end{array}
\right)      &   \left(
\begin{array}{c}
 42 
\end{array}
    \right)
 & 21       
     \\[20pt]

   \hline
  \end{tabular}
\end{table}

\newpage
\begin{table}[H]
  \centering
  \begin{tabular}{|L||P|P|L|}
    \hline
 \textbf{ CY} & \textbf{Stab} \boldsymbol{(\Phi)}  & \textbf{Modes}&  \boldsymbol{\textbf{dim}_{\mathbb H}HB}      \\
    \hline
    &&&
 \\[-15pt]
     
      (A_6,D_{8})  &  \left(
\begin{array}{cccc}
 \mathbb{Z}_2 &  &  &  \\
  &\mathbb{Z}_2  &  &  \\
  &  & \mathbb{Z}_2 &  \\
  &  &  &  \mathbb{Z}_2\\
\end{array}
\right)      &   \left(
\begin{array}{cccc}
 0 & \textcolor{ForestGreen}{4} & \textcolor{ForestGreen}{4} & \textcolor{ForestGreen}{4} \\
 \textcolor{ForestGreen}{4} & 0 & \textcolor{ForestGreen}{4} & \textcolor{ForestGreen}{4} \\
 \textcolor{ForestGreen}{4} & \textcolor{ForestGreen}{4} & 0 &\textcolor{ForestGreen}{4} \\
 \textcolor{ForestGreen}{4} & \textcolor{ForestGreen}{4} & \textcolor{ForestGreen}{4} & 0 \\
\end{array}
\right)
 & 24               \\
 &&& \\[-15pt]
    \hline
   &&& \\[-15pt]
      (A_6,D_{9})  &  \left(
\begin{array}{c}
 \mathbb{Z}_2 \\
\end{array}
\right)      &   \left(
\begin{array}{c}
 54 
\end{array}
    \right)
 & 27                \\
 &&& \\[-15pt]
    \hline
   &&& \\[-15pt]
      (A_6,D_{10})  &  \left(
\begin{array}{c}
 \mathbb{Z}_2 \\
\end{array}
\right)      &   \left(
\begin{array}{c}
 60 
\end{array}
    \right)
 & 30                \\
 &&& \\[-15pt]
    \hline
   &&& \\[-15pt]
  (A_6,D_{11})  &  \left(
\begin{array}{c}
 \mathbb{Z}_2 \\
\end{array}
\right)      &   \left(
\begin{array}{c}
 66 
\end{array}
    \right)
 & 33                \\
 &&& \\[-15pt]
    \hline
   &&& \\[-15pt]
     (A_6,D_{12})  &  \left(
\begin{array}{c}
 \mathbb{Z}_2 \\
\end{array}
\right)      &   \left(
\begin{array}{c}
72
\end{array}
    \right)
 & 36                 \\
 &&& \\[-15pt]
    \hline
   &&& \\[-15pt]

  (A_6,D_{13})  &  \left(
\begin{array}{c}
 \mathbb{Z}_2 \\
\end{array}
\right)      &   \left(
\begin{array}{c}
 78 
\end{array}
    \right)
 & 39                \\
 &&& \\[-15pt]
    \hline
   &&& \\[-15pt]
     (A_6,D_{14})  &  \left(
\begin{array}{c}
 \mathbb{Z}_2 \\
\end{array}
\right)      &   \left(
\begin{array}{c}
84
\end{array}
    \right)
 & 42                \\
 &&& \\[-15pt]
    \hline
   &&& \\[-15pt]
     (A_6,D_{15})  &  \left(
\begin{array}{cccc}
 \mathbb{Z}_2 &  &  &  \\
  &\mathbb{Z}_2  &  &  \\
  &  & \mathbb{Z}_2 &  \\
  &  &  &  \mathbb{Z}_2\\
\end{array}
\right)      &   \left(
\begin{array}{cccc}
 0 & \textcolor{ForestGreen}{6} & \textcolor{ForestGreen}{6} & \textcolor{ForestGreen}{6} \\
 \textcolor{ForestGreen}{6} & 2 & \textcolor{ForestGreen}{8} & \textcolor{ForestGreen}{8} \\
 \textcolor{ForestGreen}{6} & \textcolor{ForestGreen}{8} & 2 & \textcolor{ForestGreen}{8} \\
 \textcolor{ForestGreen}{6} & \textcolor{ForestGreen}{8} & \textcolor{ForestGreen}{8} & 2 \\
\end{array}
\right)
 & 45                \\[50pt]

   \hline
  \end{tabular}
\end{table}

\newpage
\begin{table}[H]
  \centering
  \begin{tabular}{|L||P|P|L|}
    \hline
 \textbf{ CY} & \textbf{Stab} \boldsymbol{(\Phi)}  & \textbf{Modes}&  \boldsymbol{\textbf{dim}_{\mathbb H}HB}      \\
    \hline
    &&&
 \\[-15pt]
     (A_7,D_{4})  &  \left(
\begin{array}{ccc}
 U(1)_{\alpha} &  &  \\
  & U(1)_{\beta}  &  \\
  &  &  U(1)_{\beta} \\
\end{array}
\right)      &   \left(
\begin{array}{ccc}
 0 & \textcolor{red}{4} & \textcolor{red}{4} \\
 \textcolor{red}{4} & 3 & \textcolor{blue}{4} \\
 \textcolor{red}{4} & \textcolor{blue}{4}  & 3 \\
\end{array}
\right)
 & 15                \\
&&& \\[-15pt]
    \hline
   &&& \\[-15pt]
     (A_7,D_{5})  &  
    \renewcommand{\arraystretch}{0.55}
    \scalemath{0.55}{\left(
\begin{array}{RRRRRRRRR}
 U(1)_{\alpha} &  & & &  &  &  &  &  \\
  & U(1)_{\beta} & & &  &  &  &  &  \\
  &  &U(1)_{\beta} & &  &  &  &  &  \\
  &  & &U(1)_{\gamma} &  &  &  &  &  \\
  &  & & & U(1)_{\gamma} &  &  &  &  \\
  &  & & &  &U(1)_{\delta}  &  &  &  \\
  &  & & &  &  &U(1)_{\delta} &  &  \\
  &  & & &  &  &  & U(1)_{\epsilon}&  \\
  &  & & &  &  &  &  &  U(1)_{\epsilon}\\
\end{array}
\right)}      & 
 \renewcommand{\arraystretch}{1.2} \scalemath{0.7}{\left(
\begin{array}{ccccccccc}
 0 & \textcolor{red}{1} & \textcolor{red}{1} & \textcolor{red}{1} & \textcolor{red}{1} & \textcolor{red}{1} & \textcolor{red}{1} & \textcolor{red}{1} & \textcolor{red}{1} \\
 \textcolor{red}{1} & 0 & 0 & \textcolor{red}{1} & \textcolor{red}{1} & \textcolor{red}{1} & \textcolor{red}{1} & \textcolor{red}{1} & \textcolor{red}{1} \\
 \textcolor{red}{1} & 0 & 0 & \textcolor{red}{1} & \textcolor{red}{1} & \textcolor{red}{1} & \textcolor{red}{1} & \textcolor{red}{1} & \textcolor{red}{1} \\
 \textcolor{red}{1} & 0 & 0 & 0 & 0 & \textcolor{red}{1} & \textcolor{red}{1} & \textcolor{red}{1} & \textcolor{red}{1} \\
 \textcolor{red}{1} & 0 & 0 & 0 & 0 & \textcolor{red}{1} & \textcolor{red}{1} & \textcolor{red}{1} & \textcolor{red}{1} \\
 \textcolor{red}{1} & 0 & 0 & 0 & 0 & 0 & 0 & \textcolor{red}{1} & \textcolor{red}{1} \\
 \textcolor{red}{1} & 0 & 0 & 0 & 0 & 0 & 0 & \textcolor{red}{1} & \textcolor{red}{1} \\
 \textcolor{red}{1} & 0 & 0 & 0 & 0 & 0 & 0 & 0 & 0 \\
 \textcolor{red}{1} & 0 & 0 & 0 & 0 & 0 & 0 & 0 & 0 \\
\end{array}
\right)}
 & 20                \\
&&& \\[-15pt]
    \hline
   &&& \\[-15pt]
     (A_7,D_{6})  &  \left(
\begin{array}{ccc}
 U(1)_{\alpha} &  &  \\
  & U(1)_{\beta}  &  \\
  &  &  U(1)_{\beta} \\
\end{array}
\right)      &   \left(
\begin{array}{ccc}
 0 & \textcolor{red}{4} & \textcolor{red}{4} \\
 \textcolor{red}{4} & 6 & \textcolor{blue}{8} \\
 \textcolor{red}{4} & \textcolor{blue}{8}  & 6 \\
\end{array}
\right)
 & 22               \\
&&& \\[-15pt]
    \hline
   &&& \\[-15pt]
    (A_7,D_{7})  &   \scalemath{0.7}{\left(
\begin{array}{ccccc}
 U(1)_{\alpha} &  &  &  &  \\
  &U(1)_{\beta}  &  &  &  \\
  &  &U(1)_{\beta}  &  &  \\
  &  &  & U(1)_{\gamma} &  \\
  &  &  &  & U(1)_{\gamma} \\
\end{array}
\right)}      &   \scalemath{0.7}{\left(
\begin{array}{ccccc}
 0 & \textcolor{red}{2} & \textcolor{red}{2} &\textcolor{red}{2} & \textcolor{red}{2} \\
 \textcolor{red}{2} & 1 & \textcolor{blue}{2} & \textcolor{red}{3}  & \textcolor{red}{3} \\
 \textcolor{red}{2} & \textcolor{blue}{2} & 1 & \textcolor{red}{3}  & \textcolor{red}{3}  \\
 \textcolor{red}{2} & \textcolor{red}{3}  & \textcolor{red}{3}  & 1 & \textcolor{blue}{2}  \\
 \textcolor{red}{2} & \textcolor{red}{3}  & \textcolor{red}{3}  & \textcolor{blue}{2} & 1 \\
\end{array}
\right)}
 & 26                \\
&&& \\[-15pt]
    \hline
   &&& \\[-15pt]
    (A_7,D_{8})  &  \left(
\begin{array}{ccc}
 U(1)_{\alpha} &  &  \\
  & U(1)_{\beta}  &  \\
  &  &  U(1)_{\beta} \\
\end{array}
\right)      &   \left(
\begin{array}{ccc}
 0 & \textcolor{red}{4} & \textcolor{red}{4} \\
 \textcolor{red}{4} & 9 & \textcolor{blue}{12} \\
 \textcolor{red}{4} & \textcolor{blue}{12} & 9 \\
\end{array}
\right)
 & 29           \\[50pt]

   \hline
  \end{tabular}
\end{table}

\newpage
\begin{table}[H]
  \centering
  \begin{tabular}{|L||P|P|L|}
    \hline
 \textbf{ CY} & \textbf{Stab} \boldsymbol{(\Phi)}  & \textbf{Modes}&  \boldsymbol{\textbf{dim}_{\mathbb H}HB}      \\
    \hline
    &&&
 \\[-15pt]
    (A_7,D_{9})  &  \scalemath{0.7}{\left(
\begin{array}{ccccc}
 U(1)_{\alpha} &  &  &  &  \\
  &\mathbb{Z}_2  &  &  &  \\
  &  &\mathbb{Z}_2 &  &  \\
  &  &  & \mathbb{Z}_2 &  \\
  &  &  &  & \mathbb{Z}_2 \\
\end{array}
\right)}      &   \scalemath{0.7}{\left(
\begin{array}{ccccc}
 0 & \textcolor{red}{2} & \textcolor{red}{2} & \textcolor{red}{2} & \textcolor{red}{2} \\
 \textcolor{red}{2} & 0 & \textcolor{ForestGreen}{4} & \textcolor{ForestGreen}{4} & \textcolor{ForestGreen}{4} \\
 \textcolor{red}{2} & \textcolor{ForestGreen}{4} & 0 & \textcolor{ForestGreen}{4} &\textcolor{ForestGreen}{4} \\
 \textcolor{red}{2} & \textcolor{ForestGreen}{4} & \textcolor{ForestGreen}{4} & 0 &\textcolor{ForestGreen}{4} \\
 \textcolor{red}{2} & \textcolor{ForestGreen}{4} & \textcolor{ForestGreen}{4} & \textcolor{ForestGreen}{4} & 0 \\
\end{array}
\right)}
 & 32                \\
&&& \\[-15pt]
    \hline
   &&& \\[-15pt]
    (A_7,D_{10})  &  \left(
\begin{array}{ccc}
 U(1)_{\alpha} &  &  \\
  & U(1)_{\beta}  &  \\
  &  &  U(1)_{\beta} \\
\end{array}
\right)      &   \left(
\begin{array}{ccc}
 0 & \textcolor{red}{4} & \textcolor{red}{4}  \\
 \textcolor{red}{4} & 12 & \textcolor{blue}{16}  \\
 \textcolor{red}{4}  & \textcolor{blue}{16}  & 12 \\
\end{array}
\right)
 & 36                \\
&&& \\[-15pt]
    \hline
   &&& \\[-15pt]
    (A_7,D_{11})  &  \scalemath{0.7}{\left(
\begin{array}{ccccc}
 U(1)_{\alpha} &  &  &  &  \\
  &U(1)_{\beta}  &  &  &  \\
  &  &U(1)_{\beta}  &  &  \\
  &  &  & U(1)_{\gamma} &  \\
  &  &  &  & U(1)_{\gamma} \\
\end{array}
\right)}      &  \scalemath{0.7}{\left(
\begin{array}{ccccc}
 0 & \textcolor{red}{2} & \textcolor{red}{2} &\textcolor{red}{2} & \textcolor{red}{2} \\
 \textcolor{red}{2} & 2 & \textcolor{blue}{4} & \textcolor{red}{5} & \textcolor{red}{5} \\
 \textcolor{red}{2} & \textcolor{blue}{4} & 2 & \textcolor{red}{5} & \textcolor{red}{5} \\
 \textcolor{red}{2} & \textcolor{red}{5} & \textcolor{red}{5} & 2 & \textcolor{blue}{4} \\
 \textcolor{red}{2} & \textcolor{red}{5} & \textcolor{red}{5} & \textcolor{blue}{4} & 2 \\
\end{array}
\right)}
 & 40                \\
&&& \\[-15pt]
    \hline
   &&& \\[-15pt]
    (A_7,D_{12})  &  \left(
\begin{array}{ccc}
 U(1)_{\alpha} &  &  \\
  & U(1)_{\beta}  &  \\
  &  &  U(1)_{\beta} \\
\end{array}
\right)      &  \left(
\begin{array}{ccc}
 0 & \textcolor{red}{4} & \textcolor{red}{4} \\
 \textcolor{red}{4} & 15 & \textcolor{blue}{20} \\
 \textcolor{red}{4} &  \textcolor{blue}{20} & 15 \\
\end{array}
\right)
 &43               \\
&&& \\[-15pt]
    \hline
   &&& \\[-15pt]
    (A_7,D_{13})  &  \renewcommand{\arraystretch}{0.55}
    \scalemath{0.55}{\left(
\begin{array}{RRRRRRRRR}
 U(1)_{\alpha} &  & & &  &  &  &  &  \\
  & U(1)_{\beta} & & &  &  &  &  &  \\
  &  &U(1)_{\beta} & &  &  &  &  &  \\
  &  & &U(1)_{\gamma} &  &  &  &  &  \\
  &  & & & U(1)_{\gamma} &  &  &  &  \\
  &  & & &  &U(1)_{\delta}  &  &  &  \\
  &  & & &  &  &U(1)_{\delta} &  &  \\
  &  & & &  &  &  & U(1)_{\epsilon}&  \\
  &  & & &  &  &  &  &  U(1)_{\epsilon}\\
\end{array}
\right)}      &   \renewcommand{\arraystretch}{1.2} \scalemath{0.7}{\left(
\begin{array}{ccccccccc}
 0 & \textcolor{red}{1} & \textcolor{red}{1} & \textcolor{red}{1}& \textcolor{red}{1} & \textcolor{red}{1} & \textcolor{red}{1} & \textcolor{red}{1} & \textcolor{red}{1} \\
 \textcolor{red}{1} & 0 & \textcolor{blue}{1} & \textcolor{red}{3} & \textcolor{red}{3}  & \textcolor{red}{3}  & \textcolor{red}{3}  & \textcolor{red}{3}  & \textcolor{red}{3}  \\
 \textcolor{red}{1} & \textcolor{blue}{1} & 0 & \textcolor{red}{3}  & \textcolor{red}{3}  & \textcolor{red}{3}  & \textcolor{red}{3}  & \textcolor{red}{3}  & \textcolor{red}{3}  \\
 \textcolor{red}{1} & 0 & 0 & 0 & \textcolor{blue}{1} & \textcolor{red}{3}  & \textcolor{red}{3}  & \textcolor{red}{3}  &\textcolor{red}{3}  \\
 \textcolor{red}{1} & 0 & 0 &\textcolor{blue}{1} & 0 & \textcolor{red}{3}  & \textcolor{red}{3}  & \textcolor{red}{3}  & \textcolor{red}{3} \\
 \textcolor{red}{1} & 0 & 0 & 0 & 0 & 0 & \textcolor{blue}{1} & \textcolor{red}{3}  & \textcolor{red}{3}  \\
 \textcolor{red}{1} & 0 & 0 & 0 & 0 & \textcolor{blue}{1} & 0 & \textcolor{red}{3}  & \textcolor{red}{3}  \\
 \textcolor{red}{1} & 0 & 0 & 0 & 0 & 0 & 0 & 0 & \textcolor{blue}{1} \\
 \textcolor{red}{1} & 0 & 0 & 0 & 0 & 0 & 0 & \textcolor{blue}{1} & 0 \\
\end{array}
\right)}
 & 48                 \\[80pt]

   \hline
  \end{tabular}
\end{table}

\newpage
\begin{table}[H]
  \centering
  \begin{tabular}{|L||P|P|L|}
    \hline
 \textbf{ CY} & \textbf{Stab} \boldsymbol{(\Phi)}  & \textbf{Modes}&  \boldsymbol{\textbf{dim}_{\mathbb H}HB}      \\
    \hline
    &&&
 \\[-15pt]
    (A_7,D_{14})  &  \left(
\begin{array}{ccc}
 U(1)_{\alpha} &  &  \\
  & U(1)_{\beta}  &  \\
  &  &  U(1)_{\beta} \\
\end{array}
\right)      &   \left(
\begin{array}{ccc}
 0 & \textcolor{red}{4} & \textcolor{red}{4} \\
 \textcolor{red}{4} & 18 & \textcolor{blue}{24} \\
 \textcolor{red}{4} &  \textcolor{blue}{24} & 18 \\
\end{array}
\right)
 & 50                \\
&&& \\[-15pt]
    \hline
   &&& \\[-15pt]
       (A_7,D_{15})  &  \scalemath{0.7}{\left(
\begin{array}{ccccc}
 U(1)_{\alpha} &  &  &  &  \\
  &U(1)_{\beta}  &  &  &  \\
  &  &U(1)_{\beta}  &  &  \\
  &  &  & U(1)_{\gamma} &  \\
  &  &  &  & U(1)_{\gamma} \\
\end{array}
\right)}      &  \scalemath{0.7}{\left(
\begin{array}{ccccc}
 0 & \textcolor{red}{2} & \textcolor{red}{2} & \textcolor{red}{2} & \textcolor{red}{2} \\
\textcolor{red}{2} & 3 & \textcolor{blue}{6} & \textcolor{red}{7} & \textcolor{red}{7}  \\
 \textcolor{red}{2} & \textcolor{blue}{6} & 3 &  \textcolor{red}{7}  &  \textcolor{red}{7}  \\
 \textcolor{red}{2} &  \textcolor{red}{7}  &  \textcolor{red}{7}  & 3 &\textcolor{blue}{6} \\
 \textcolor{red}{2} &  \textcolor{red}{7}  &  \textcolor{red}{7}  & \textcolor{blue}{6} & 3 \\
\end{array}
\right)}
 & 54                \\[50pt]

   \hline
  \end{tabular}
\end{table}

\begin{table}[H]
  \centering
  \begin{tabular}{|L||P|P|L|}
    \hline
 \textbf{ CY} & \textbf{Stab} \boldsymbol{(\Phi)}  & \textbf{Modes}&  \boldsymbol{\textbf{dim}_{\mathbb H}HB}      \\
    \hline
    &&&
 \\[-15pt]
     (A_8,D_{4})  &  \left(
\begin{array}{cc}
 \mathbb{Z}_2 &  \\
  & \mathbb{Z}_2 \\
\end{array}
\right)      &   \left(
\begin{array}{cc}
 4 & \textcolor{ForestGreen}{12} \\
 \textcolor{ForestGreen}{12} & 4 \\
\end{array}
\right)
 & 16                \\
&&& \\[-15pt]
    \hline
   &&& \\[-15pt]
       (A_8,D_{5})  &  \left(
\begin{array}{c}
 \mathbb{Z}_2 \\
\end{array}
\right)      &   \left(
\begin{array}{c}
 40 
\end{array}
    \right)
 & 20                \\
&&& \\[-15pt]
    \hline
   &&& \\[-15pt]
       (A_8,D_{6})  &  \left(
\begin{array}{c}
 \mathbb{Z}_2 \\
\end{array}
\right)      &   \left(
\begin{array}{c}
 48 
\end{array}
    \right)
 & 24                \\
&&& \\[-15pt]
    \hline
   &&& \\[-15pt]
       (A_8,D_{7})  &  \left(
\begin{array}{cc}
 \mathbb{Z}_2 &  \\
  & \mathbb{Z}_2 \\
\end{array}
\right)      &   \left(
\begin{array}{cc}
 6 & \textcolor{ForestGreen}{18} \\
 \textcolor{ForestGreen}{18} & 14 \\
\end{array}
\right)
 & 28                \\[30pt]
   
   \hline
  \end{tabular}
\end{table} 

\newpage
\begin{table}[H]
  \centering
  \begin{tabular}{|L||P|P|L|}
    \hline
 \textbf{ CY} & \textbf{Stab} \boldsymbol{(\Phi)}  & \textbf{Modes}&  \boldsymbol{\textbf{dim}_{\mathbb H}HB}      \\
    \hline
    &&&
 \\[-15pt]
       (A_8,D_{8})  &  \left(
\begin{array}{c}
 \mathbb{Z}_2 \\
\end{array}
\right)      &   \left(
\begin{array}{c}
 64 
\end{array}
    \right)
 & 32                \\
&&& \\[-15pt]
    \hline
   &&& \\[-15pt]
       (A_8,D_{9})  &  \left(
\begin{array}{c}
 \mathbb{Z}_2 \\
\end{array}
\right)      &   \left(
\begin{array}{c}
 72 
\end{array}
    \right)
 & 36                \\
&&& \\[-15pt]
    \hline
   &&& \\[-15pt]
       (A_8,D_{10})  &  \left(
\begin{array}{ccccc}
\mathbb{Z}_2  &  &  &  &  \\
  &\mathbb{Z}_2  &  &  &  \\
  &  & \mathbb{Z}_2 &  &  \\
  &  &  & \mathbb{Z}_2 &  \\
  &  &  &  &  \mathbb{Z}_2\\
\end{array}
\right)      &  \left(
\begin{array}{ccccc}
 0 & \textcolor{ForestGreen}{4} & \textcolor{ForestGreen}{4} & \textcolor{ForestGreen}{4} & \textcolor{ForestGreen}{4} \\
 \textcolor{ForestGreen}{4} & 0 & \textcolor{ForestGreen}{4} & \textcolor{ForestGreen}{4} & \textcolor{ForestGreen}{4} \\
 \textcolor{ForestGreen}{4} & \textcolor{ForestGreen}{4} & 0 & \textcolor{ForestGreen}{4} & \textcolor{ForestGreen}{4} \\
 \textcolor{ForestGreen}{4} &\textcolor{ForestGreen}{4} & \textcolor{ForestGreen}{4} & 0 & \textcolor{ForestGreen}{4} \\
 \textcolor{ForestGreen}{4} & \textcolor{ForestGreen}{4} & \textcolor{ForestGreen}{4} & \textcolor{ForestGreen}{4} & 0 \\
\end{array}
\right)
 & 40                \\
&&& \\[-15pt]
    \hline
   &&& \\[-15pt]
       (A_8,D_{11})  &  \left(
\begin{array}{c}
 \mathbb{Z}_2 \\
\end{array}
\right)      &   \left(
\begin{array}{c}
 88 
\end{array}
    \right)
 & 44                \\
&&& \\[-15pt]
    \hline
   &&& \\[-15pt]
     (A_8,D_{12})  &  \left(
\begin{array}{c}
 \mathbb{Z}_2 \\
\end{array}
\right)      &   \left(
\begin{array}{c}
 96 
\end{array}
    \right)
 & 48                \\
&&& \\[-15pt]
    \hline
   &&& \\[-15pt]
     (A_8,D_{13})  &  \left(
\begin{array}{cc}
 \mathbb{Z}_2 &  \\
  & \mathbb{Z}_2 \\
\end{array}
\right)      &   \left(
\begin{array}{cc}
 10 & \textcolor{ForestGreen}{30} \\
 \textcolor{ForestGreen}{30} & 34 \\
\end{array}
\right)
 & 52                \\
&&& \\[-15pt]
    \hline
   &&& \\[-15pt]
     (A_8,D_{14})  &  \left(
\begin{array}{c}
 \mathbb{Z}_2 \\
\end{array}
\right)      &   \left(
\begin{array}{c}
 112 
\end{array}
    \right)
 & 56                \\
&&& \\[-15pt]
    \hline
   &&& \\[-15pt]
       (A_8,D_{15})  &  \left(
\begin{array}{c}
 \mathbb{Z}_2 \\
\end{array}
\right)      &   \left(
\begin{array}{c}
 120 
\end{array}
    \right)
 & 60                \\[20pt]

   \hline
  \end{tabular}
\end{table}

\restoregeometry

\end{document}